\documentclass[showpacs,pre,twocolumn,floatfix]{revtex4}
\usepackage{graphicx}
\usepackage{amsmath}
\usepackage{dcolumn}

\begin{document}

\title{Monte Carlo study of the evaporation/condensation transition on
different Ising lattices}

\author{A. Nu{\ss}baumer}
\author{E. Bittner}
\author{W. Janke}

\affiliation{
Institut f\"ur Theoretische Physik and Centre for Theoretical
Sciences (NTZ) -- Universit\"at Leipzig, Postfach 100\,920, D-04009
Leipzig, Germany\\
}


\newcommand{\etal}{\unskip\ \emph{et al.}}
\newcommand{\upd}{\rm d}

\begin{abstract}
In 2002 Biskup \etal{} [{\it Europhys.\ Lett.} {\bf 60}, 21 (2002)] sketched
a rigorous proof for the behavior of the 2D Ising lattice gas, which is
equivalent to the ordinary spin-$1/2$ Ising model, at a finite volume and a
fixed excess $\delta M$ of particles (spins) above the ambient gas density
(spontaneous magnetisation). By identifying a dimensionless parameter $\Delta
(\delta M)$ and a universal constant $\Delta_\mathrm{c}$, they showed in the
limit of large system sizes that for $\Delta < \Delta_c$ the excess is absorbed
in the background (``evaporated'' system), while for $\Delta > \Delta_c$ a
droplet of the dense phase occurs (``condensed'' system). By minimising the free
energy of the system they derive an explicit formula for the fraction
$\lambda(\Delta)$ of excess particles forming the droplet.

To check the applicability of the analytical results to much smaller,
practically accessible system sizes, we performed several Monte Carlo
simulations for the 2D Ising model with nearest-neighbour couplings on a square
lattice at fixed magnetisation $M$. Thereby, we measured the largest minority
droplet, corresponding to the condensed phase, at various system sizes ($L=40,
\dots, 640$). With analytic values for for the spontaneous magnetisation $m_0$,
the susceptibility $\chi$ and the Wulff interfacial free energy density
$\tau_\mathrm{W}$ for the infinite system, we were able to determine $\lambda$
numerically in very good agreement with the theoretical prediction.

Furthermore, we did simulations for the spin-1/2 Ising model on a triangular
lattice and with next-nearest-neighbour couplings on a square lattice. Again,
finding a very good agreement with the analytic formula, we demonstrate the
universal aspects of the theory with respect to the underlying lattice. For
the case of the next-nearest-neighbour model, where $\tau_\text W$ is unknown
analytically, we present different methods to obtain it numerically by fitting
to the distribution of the magnetisation density $P(m)$.

\end{abstract}

\maketitle

\section{Introduction}

\begin{figure}
 \begin{center}
  \includegraphics[scale=0.7]{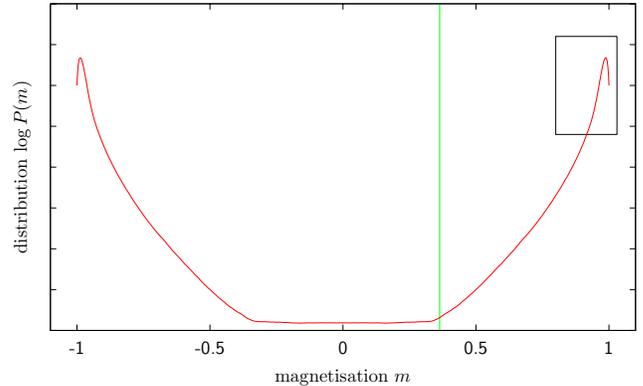}
  \caption{Schematic plot of the probability density $P(m)$ of
  the magnetisation in logarithmic form. The marked box indicates the
  position of the cut-out displayed in Fig.~\ref{fig:cusp}. The vertical
  (green) line indicates the droplet/strip transition point for positive
magnetisation $m>0$, the use of  which will be explained later on in
Sec.~\ref{sec:parameters.nnn}.}
  \label{fig:cusp.schematic}
 \end{center}
\end{figure}

\begin{figure}
 \begin{center}
  \includegraphics[scale=0.7]{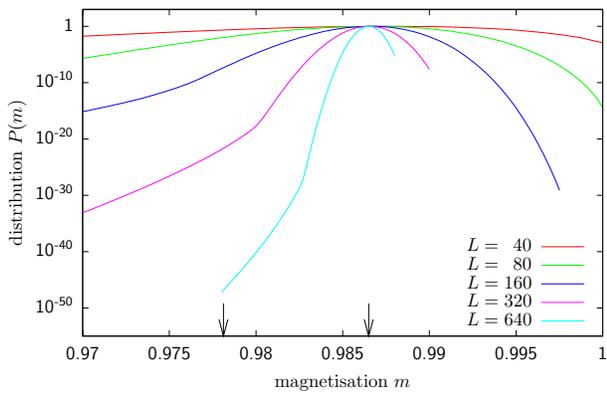}
  \caption{Probability density of the magnetisation for the
  two-dimensional Ising model around its right peak for different
  system sizes $L$ at the temperature $T=1.5$. The cusp indicates the
  evaporation/condensation transition region.  On the right side of the
  cusp (evaporated system) a Gaussian peak is clearly visible, while on
  the left side (condensed system) a stretched exponential
  behavior can be seen. The two arrows on the $x$-axis indicate for
  $L=640$ the range of data points shown in Fig.~\ref{fig:nn.lambda.delta}.}
  \label{fig:cusp}
 \end{center}
\end{figure}

\begin{figure}
  \includegraphics[scale=0.35]{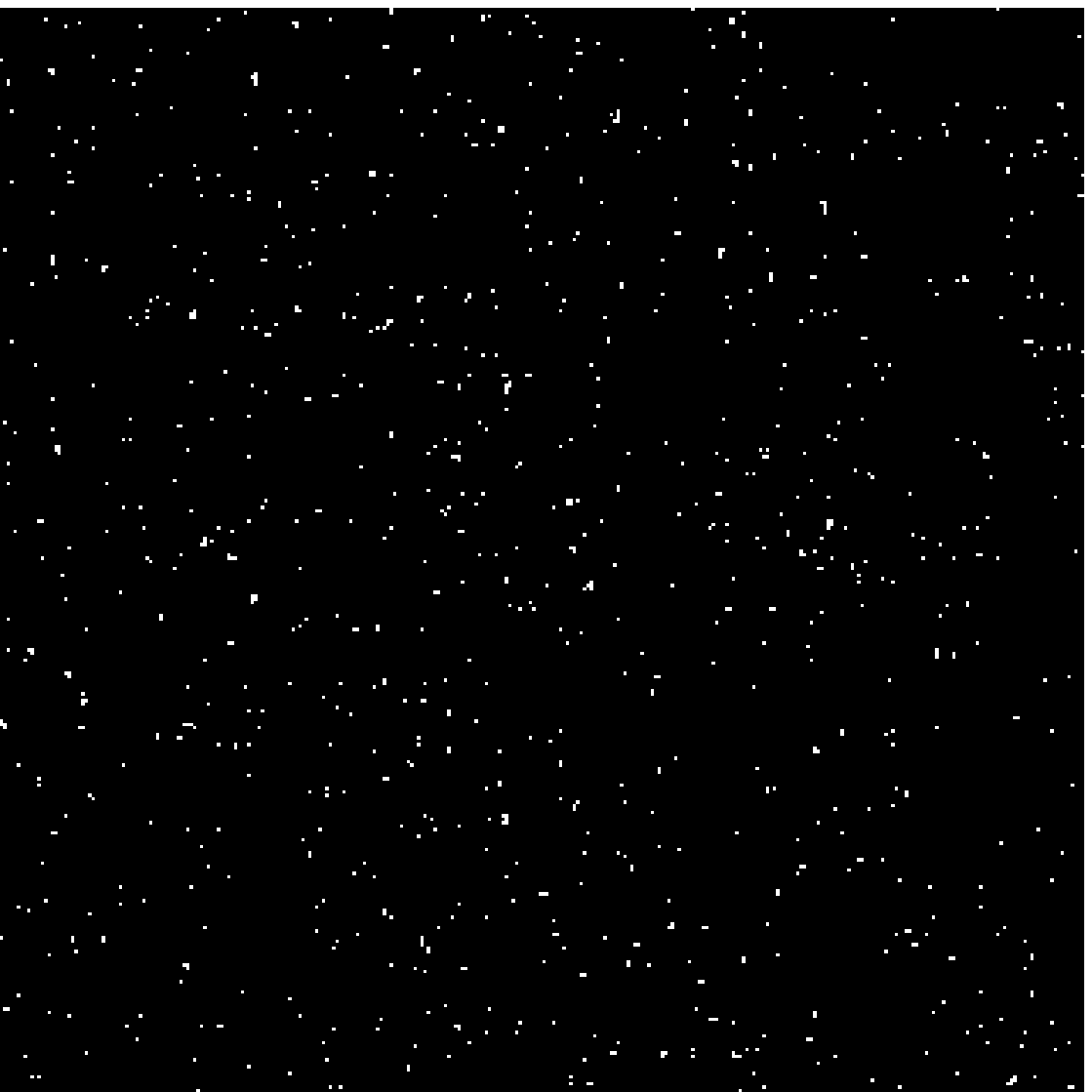}
  \hspace{0.5cm}
  \includegraphics[scale=0.35]{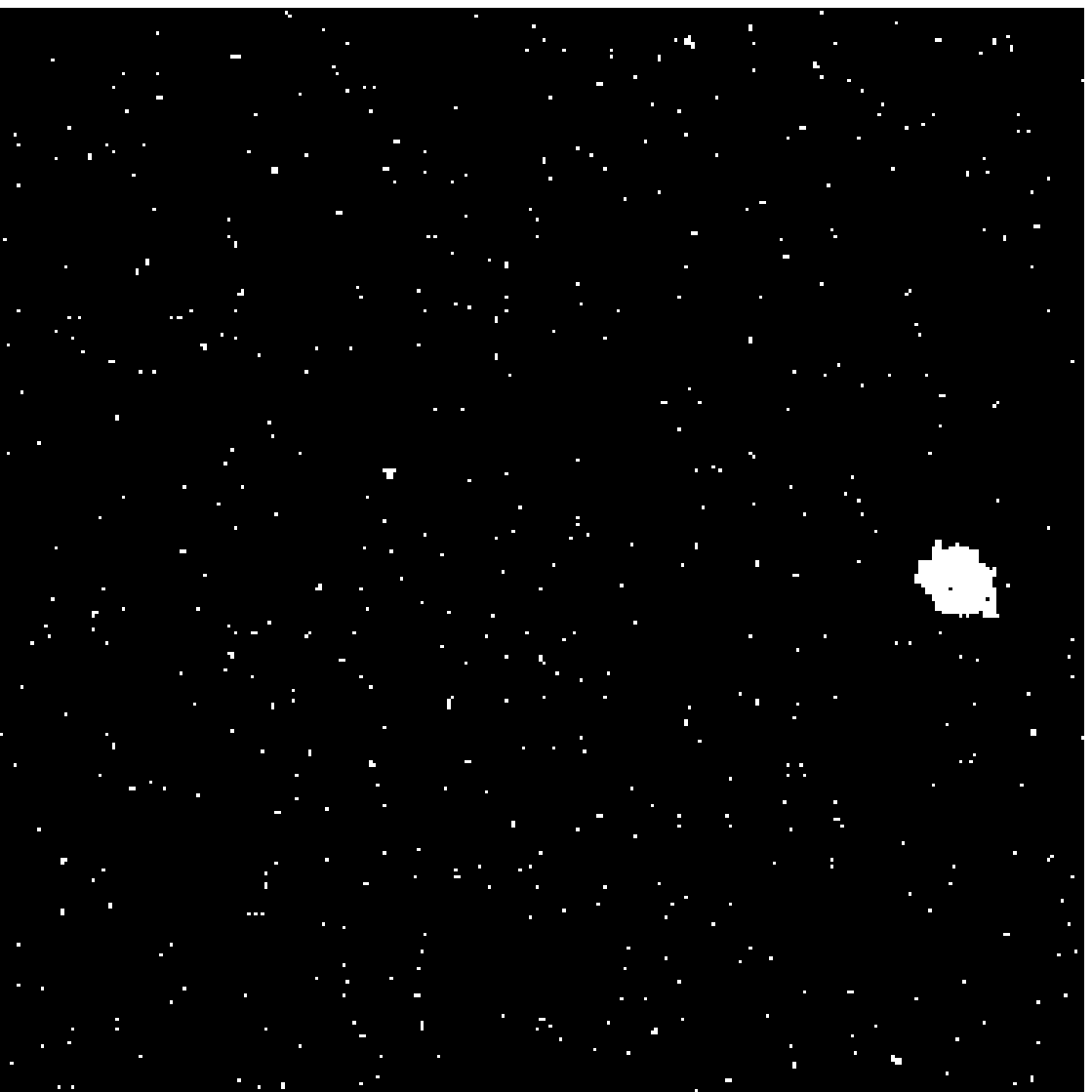}
  \caption{Two snapshots of a $320 \times 320$ n.n.\ Ising system at
$T=1.5$ and the same value of the magnetisation $m=0.9801$ chosen to be in the
vicinity of the evaporation/condensation point. Left: Evaporated system, a large
number of very small excitations (bubbles) exist (1 to 4 spins) and the largest
cluster consists of 5 connected spins. Right: Condensed system, a single large
droplet with volume 400 spins that has absorbed a large amount of the small
bubbles.}
  \label{fig:ce}
\end{figure}

The formation and dissolution of equilibrium droplets at a first-order phase
transition is one of the longstanding problems in statistical mechanics
\cite{fisher}. Quantities of particular interest are the size and free energy of
a ``critical droplet'' that needs to be formed before the decay of the
metastable state via homogeneous nucleation can start. For large but finite
systems, this is signalised by a cusp in the probability density of the order
parameter $\phi$ towards the phase-coexistence region as depicted in
Figs.~\ref{fig:cusp.schematic} and \ref{fig:cusp} for the example of the
two-dimensional (2D) Ising model, where $\phi=m$ is the magnetisation. This
evaporation/condensation ``transition point'' separates an ``evaporated'' phase
with many very small bubbles of the ``wrong'' phase around the peak at $\phi_0$
from the ``condensed phase'' phase, in which a large droplet has formed; for
configuration snapshots see Fig.~\ref{fig:ce}. The droplet eventually grows
further towards $\phi=0$ until it percolates the finite system in another
droplet/strip ``transition''. The latter transition is indicated in the 2D Ising
model by the cusp at the beginning of the flat two-phase region around $m=0$
(see Fig.~\ref{fig:cusp.schematic}).

Building on the seminal work by Fisher \cite{fisher} developing the droplet
picture, early numerical studies of the evaporation/condensation transition by
Binder, Kalos and Furukawa \cite{binder.kalos,furukawa.binder} date back to the
beginning of the 1980s.  Recently this problem has been taken up again by
Neuhaus and Hager \cite{neuhaus.hager} who discussed it with emphasis on
possible Gibbs-Thomson and Tolman corrections. This stimulated further new
theoretical \cite{biskup,biskup2,binder} and numerical
\cite{virnau.binder,nussbaumer:06} work.

Here, we follow the exposition of Biskup \etal{} \cite{biskup,biskup2}, who
present their results both in a phenomenological liquid-vapour (or solid-gas)
picture and also explicitly in terms of the simple Ising (lattice-gas) model. 
The distinguishing feature of their work is the formulation of a proper
equilibrium theory which does not 
need to explicitly involve correction effects a la Gibbs-Thomson or Tolman
\cite{biskup3} as was done in earlier works
\cite{lee:95,pleimling:00,pleimling:01}. We consider this feature as one of the
main merits of their formulation which can be shown to be equivalent (at least
in leading order) to the earlier less rigorous treatment in
\cite{neuhaus.hager}.

The price one has to pay, however, is a rather intricate rescaling of the
original problem which requires in numerical work great care with details. To
set the theoretical grounds for our Monte Carlo simulation study and in
particular to develop intuition for the final representation of our results in
Figs.~\ref{fig:nn.lambda.delta}--\ref{fig:lambda.delta.640}, we therefore start
first with a brief summary of the Biskup \etal{} \cite{biskup,biskup2} theory.
In order to do so, we restrict ourselves to the special case of the 2D Ising model
with Hamiltonian
\begin{equation}
 {\cal H} = -J \sum_{\left< i,j\right>} s_i s_j \,,
 \label{eq:ising}
\end{equation}
where $s_i=\pm 1$ and $\left<i,j \right>$ denotes a (next-)nearest-neighbour
pair. If a down-spin ($\sigma_i=-1$) is treated as a particle and an up-spin
($\sigma_i=1$) as a vacancy, the system can be interpreted as a lattice gas of
atoms.

%


\section{Theory\label{sec:theory}}
In this section we summarise the considerations of Biskup \etal{} \cite{biskup}
but specialised for the case of the two-dimensional Ising model (not
necessarily on a square lattice).

\begin{figure}
 \includegraphics{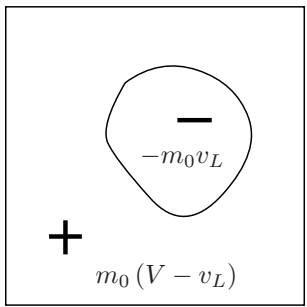}
 \caption{Ising system of size $V$ with a minority droplet of volume
  $v_L$ of negative spins surrounded by positive spins with a volume
  $(V-v_L)$, shown in the extreme case where the total excess
  in magnetisation is concentrated in the droplet, i.e. $v_\text d=v_L$.}
  \label{fig:ising}
\end{figure}

We image the following situation: an {\it unconstrained\/} \footnote{Later on we
will mostly look at constrained systems, i.e.\ the magnetisation is fixed.}
Ising system of size $V=L\times L$ in the low-temperature phase at the inverse
temperature $\beta \equiv J / k_\mathrm B T > \beta_\mathrm c$. If the majority
of spins is positive ($\sigma_i = 1$), i.e., the system is in the phase with
positive magnetisation, then, due to thermal fluctuations, there are always
some overturned negative spins and the total magnetisation is $M=m_0 V$, with
$m_0 < 1$. Here, $m_0 = m_0(\beta) > 0$ denotes the infinite-volume equilibrium
magnetisation (spontaneous magnetisation) as, e.g., calculated analytically by
Onsager and Yang for the square lattice with next-neighbour interactions (see
Sec.~\ref{sec:setup}). Now, if some volume $v_L$ of the systems is inverted
\footnote{Inversion of spins means the operation $\sigma_i \to - \sigma_i$
for \textit{all} spins in the volume $v_L$.}, then the magnetisation of this
{\it constrained} system is
\begin{equation}
 M = m_0(V-v_L)-m_0 v_L \;.
 \label{eq:M.vL}
\end{equation}
It is important to note, that here we did not require the inverted volume $v_L$
to be connected or to be of the form of a droplet. Still, we present in
Fig.~\ref{fig:ising} this extreme case to make it simpler to identify the
quantities introduced here. Secondly, as only spins
are inverted, by symmetry it must hold exactly $m_0^{(-)} = -m_0^{(+)}$,
otherwise, an
completely inverted system would have another value for the spontaneous
magnetisation.
Now, the difference to the original, unconstrained system with magnetisation
$M_0=m_0 V$ is
\begin{equation}
 \delta M = M - M_0 = -2 v_L m_0 \;.
 \label{eq:deltaM.vL}
\end{equation}
The factor $2$ is due to the definition of the Ising spins, having a value $\pm
1$. The interpretation of this formula is as follows: a system which has
a difference in the magnetisation of $\delta M$ to an unconstrained Ising
system has a volume $v_L$ of inverted spins. Biskup \etal{} show that for a
given magnetisation $M$ the total volume of inverted spins $v_L$ can be (in the
thermodynamic limit of large systems) divided into two parts, unconnected
small fluctuations with volume $v_\text f$ and a single large connected droplet
with volume $v_\text d$, since there exist no droplets of intermediate size
\cite{biskup2}. For the total volume of inverted spins holds $v_L = v_\text f +
v_\text d$.

Now, the free energy can be decomposed according to the two contributions. For
the droplet it is written as
\begin{equation}
 F_d = \tau_\mathrm{W} \sqrt{v_\text d}\;,
 \label{eq:F.d}
\end{equation}
where $\tau_\mathrm W$ is the interfacial free energy per unit volume of an
ideally shaped droplet, also known as the free energy of a droplet of Wulff
shape \cite{wulff}. The contribution of the fluctuations is derived in the
following manner. From the volume $V$ of the whole system already $v_\text d$ is
occupied by the single large droplet. The rest of the system has an
unconstrained magnetisation of $M_0^f =(V-v_\text d) m_0$. If some volume
$v_\text f$ of the
remaining spins is inverted, then the magnetisation is
\begin{equation}
 M^\text f=(V-v_\text d-v_\text f) m_0 -m_0 v_\text f \;.
\end{equation}
Then, the difference $\delta M^\text f$ to the unconstrained magnetisation
$M_0^f$ is
\begin{equation}
 \delta M^\text f = M^\text f - M_0^\text f = -2 m_0 v_\text f \;.
\end{equation}
The contribution to the free energy due to these fluctuations can be written as
\begin{equation}
 F_\text f = \frac{\left(M^\text f-M_0^\text f\right)^2}{2 \chi V}
     = \frac{2 m_0^2 v_\text f^2}{\chi V} \;,
 \label{eq:F.f}
\end{equation}
where $\chi = \chi(\beta)=\beta V [\langle m^2\rangle- \langle m\rangle^2
]$ is the susceptibility in the thermodynamic limit.

Now, the relative volume of
the droplet compared to the total volume of overturned spins $v_L$ is defined as
\begin{equation}
 \lambda = \frac{v_\text d}{v_L}
 \hspace{1cm}\text{or}\hspace{1cm}
 v_\text d=\lambda v_L \;.
 \label{eq:lambda}
\end{equation}
Hence, $v_\text f$ can be written as
\begin{equation}
 v_\text f = v_L-v_\text d = v_L\left(1-\frac{v_\text d}{v_L} \right) = v_L
\left(1-\lambda \right)
\;.
\end{equation}
Using this relation, the total free energy $F= F_\text d + F_\text f $ is
\begin{align}
 F = & \tau_\mathrm{W} \sqrt{v_\text d} + \frac{2 m_0^2 v_\text f^2}{\chi V} \\
   = & \tau_\mathrm{W} \sqrt{\lambda v_L} + \frac{2 m_0^2}{\chi V}
v_L^2(1-\lambda)^2 \;,
 \label{eq:F.total}
\end{align}
or, in the form of Biskup \etal{}, 
\begin{equation}
 F(\lambda) = \tau_\mathrm{W} \sqrt{v_L} \phi_\Delta (\lambda)
 \label{eq:F.Kotecky}
\end{equation}
with
\begin{equation}
 \phi_\Delta(\lambda) = \sqrt{\lambda} + \Delta \left( 1-\lambda \right)^2 \;,
\end{equation}
and
\begin{equation}
 \Delta = \frac{2 m_0^2 v_L^2}{\chi V \tau_\mathrm{W}
\sqrt{v_L}}=\frac{2 m_0^2 v_L^{3/2}}{\chi V \tau_\mathrm{W}}  \;.
\label{eq:Delta}
\end{equation}
Now, if the magnetisation is fixed to some value, then the total number of
overturned spins is also fixed and using Eq.~(\ref{eq:M.vL}) it holds
\begin{equation}
 v_L = \frac 12 \left(V - \frac M {m_0} \right) \;.
\end{equation}
As $m_0$, $\chi$ and $\tau_\mathrm W$ are constants, the only varying quantity
in Eq.~(\ref{eq:F.Kotecky}) is the relative volume of the droplet $\lambda$. A
fully equilibrated thermodynamic system always stays in the minimum of the free
energy.  Therefore, the physical $\lambda_\Delta$, i.e., the correct
distribution of overturned volume between the droplet and the fluctuations,
minimises $F$ in the range $\lambda \in [0,1]$.  Consequently, the solution of
this problem is either given by $\frac{\partial \phi_\Delta}{\partial
\lambda}=0$, which is
\begin{equation}
 \frac 1 { 2 \sqrt \lambda} - 2 \Delta  (1-\lambda)=0 \;,
 \label{eq:F.min}
\end{equation}
or it is one of the boundary values $0,1$. Solving  Eq.~(\ref{eq:F.min}) shows
that for $\Delta < \Delta_\mathrm c$ the correct solution  is $\lambda  =
0$, i.e., pure fluctuations and no droplet at all. The point $\Delta_\mathrm c$
it given by the condition $\phi_{\Delta_\text c}(0) =
\phi_{\Delta_\text c}(\lambda_\text c)$ which is $\Delta_\text c =
\sqrt{\lambda_\text c} + \Delta_\text c (1-\lambda_\text c)^2$ or
\begin{equation}
 \Delta_{\mathrm c} = \frac 1 {\sqrt{\lambda_\mathrm c} (2-\lambda_{\mathrm c})}
\;.
 \label{eq:Deltac.lambdac}
\end{equation}
This can be substituted in Eq.~(\ref{eq:F.min}) resulting in $\frac 1 {2
\sqrt{\lambda_\mathrm c}} -\frac{2 (1-\lambda_{\mathrm
c})}{\sqrt{\lambda_\mathrm c}(2-\lambda_{\mathrm c})} = 0$ or
\begin{equation}
 \lambda_{\mathrm c} = \frac 23 \;.
\end{equation}
Inserting this value into Eq.~(\ref{eq:Deltac.lambdac}) gives 
\begin{equation}
 \Delta_\text c = \frac 3 4 \sqrt \frac 3 2 = 0.918558 \dots \;.
\end{equation}
For $\Delta > \Delta_{\mathrm c}$ the solution is 
\begin{equation}
 \lambda = \frac 4 3 \cos^2\left[
 \frac{\pi - \cos^{-1}\left( \frac{3 \sqrt 3}{8 \Delta} \right)}{3}\right] \;.
 \label{eq:lambda.nontrivial}
\end{equation}

\begin{figure}
  \includegraphics[scale=0.7]{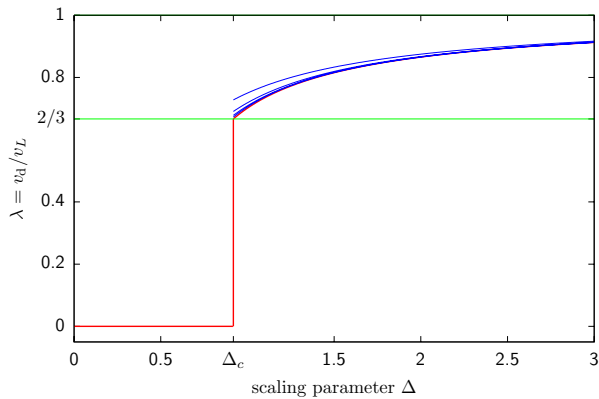}
  \caption{Fraction of the excess magnetisation in the largest droplet
$\lambda$ in dependence of the scaling parameter $\Delta$. For
$\Delta<\Delta_\text c$ there is no largest droplet, only fluctuations. At
$\Delta=\Delta_\text c$ a droplet is formed, containing $2/3$ of the total
excess. In the case $\Delta > \Delta_\text c$ the fraction of the excess is
given by Eq.~(\ref{eq:lambda.nontrivial}). The (blue) lines approaching
$\lambda$ for $\Delta > \Delta_\text c$ are the Taylor series of
Eq.~(\ref{eq:lambda.nontrivial}) up to order 4 around $\Delta = \infty$ that
have the form $\lambda =
1-1/4\Delta - 1/32 \Delta^2 -5/512 \Delta^3 - 1/256 \Delta^4 + \dots$ .}
\label{fig:lambda.delta.analytic}
\end{figure}

These results give rise to the following physical picture. For fixed
magnetisation $M \approx M_0$, where $\Delta(M)<\Delta_\mathrm c$, the systems
contains no droplet, only fluctuations are present. At some value $M_\mathrm c$
with $\Delta(M_\mathrm c) = \Delta_\mathrm c$ two states coexist, the state
of pure fluctuations and a mixed state composed of a droplet that absorbs
$2/3$  of the fluctuations and the remaining $1/3$ of the fluctuations. For
smaller magnetisation, i.e.\ $\Delta(M) > \Delta_\text c$, the droplet grows and
thereby absorbs more and more of the background fluctuations. The predicted
behavior of $\lambda = \lambda(\Delta)$ is shown in
Fig.~\ref{fig:lambda.delta.analytic}.

\section{Set up}
\label{sec:setup}

In this work we wanted to answer two questions. On the one hand, we wanted to
test from which system sizes on the theoretical results presented in the last
section start to yield a good description of the data for the two-dimensional
Ising model. On the other hand, we wanted to check the universal aspects of the
theory by using different lattice models, namely the triangular
nearest-neighbour (n.n.) lattice and the next-nearest neighbour (n.n.n.) square
lattice. In order to do so, $\lambda$, the fraction of the excess
of magnetisation in the largest droplet defined in Eq.~(\ref{eq:lambda}), had to
be measured in dependence of the parameter $\Delta$ defined in
Eq.~(\ref{eq:Delta}).

To get the correct scaling for the abscissa, the parameter $\Delta(v_L,m_0,
\chi,\tau_\mathrm{W})$ had to be calculated according to Eq.~(\ref{eq:Delta}).
While $v_L$ is a free parameter, the magnetisation, the susceptibility and the
free energy of the Wulff droplet per unit volume must be obtained analytically
or by other means, e.g., as results of simulations. For the free energy
of the Wulff droplet the analytic expression $\Sigma_\mathrm{W} = 2 \sqrt{W
\Sigma}$, e.g.\ \cite{zia:82,leung:90}, can be used. Here,
$\Sigma$ is the volume of the droplet and $W$ is the volume bounded by the Wulff
plot. Putting $\Sigma=1$ gives the interfacial free energy per unit volume
\begin{equation}
  \tau_\mathrm{W}(\beta)=2 \sqrt{W} \,.
  \label{eq:interface.tension}
\end{equation}

In the following three subsections we discuss for the three studied models the
origin of the constants in question. For the standard Ising model
with nearest-neighbour couplings on a square lattice and the Ising model on a
triangular lattice all relevant constants are known from literature, either
analytically or from quite long series expansions. This is not the case,
however, for the n.n.n.\ Ising model and, therefore, here we had to apply
simulations to retrieve the values.

\subsection{Parameters for the n.n.\ Ising model on a square lattice}

The critical temperature of the Ising model was given in 1941 by Kramers and
Wannier \cite{kramers:41}. Using self-duality arguments they obtained the
expression
\begin{equation}
 T_\mathrm c = \frac 2 { \ln(1+\sqrt 2)} \;.
\end{equation}
For the spontaneous magnetisation $m_0$ there exists the famous Onsager-Yang
analytic solution \cite{onsager, yang}
\begin{equation} 
  m_0(\beta)=\left[1-\sinh^{-4}\left(2 \beta \right)\right]^{1/8} \,.
  \label{eq:nn.m0} 
\end{equation}
Also the susceptibility $\chi$ is virtually known to arbitrary precision
from very long series expansions, e.g., Orrick \etal{} \cite{orrick} give the
formula
\begin{equation}
  \chi(\beta) = \beta \sum_{i=0}^n c_i u^{2 i} 
  \hspace{0.5cm}\text{with}\hspace{0.5cm}
  u = \frac{1}{2 \sinh(2 \beta)} 
  \label{eq:nn.chi}
\end{equation}
and $c = \{$0, 0, 4, 16, 104, 416, 2\,224, 8\,896, 43\,840, 175\,296, 825\,648,
3\,300\,480, 15\,101\,920, $...\}$ up to order $323$ (at $T=1.5$ the last term
contributes $\approx 0.28 \times 10^{-158}$).
The volume of the Wulff plot is given by \cite{leung:90}
\begin{equation}
  W = \frac{4}{\beta^2}\int_0^{\beta \sigma_0} \upd x 
      \cosh^{-1} \left[ \frac{\cosh^2(2\beta)}{\sinh(2\beta)} 
      -\cosh(x)\right] \,,
\end{equation}
where
\begin{equation}
 \sigma_0 = 2 + \frac 1 \beta \ln [\tanh(\beta)]
 \label{eq:nn:simga0}
\end{equation}
is the interface tension of the (1,0) surface (i.e., in direction of the axis).
For the (1,1) surface the exact expressions reads \cite{fisher:67,rottman:81}
\begin{equation}
 \sigma_1 = \frac{\sqrt 2}\beta \ln \left[ \sinh \left( 2
\beta \right) \right] \;.
 \label{eq:nn.sigma1}
\end{equation}

\subsection{Parameters for the n.n.\ Ising model on a triangular lattice}

The critical temperature of the triangular lattice is \cite{baxter:82}
\begin{equation}
 T_\text c = \frac 4 {\ln 3} \;.
\end{equation}
For the spontaneous magnetisation Potts \cite{potts.m0} gave in 1952
the expression
\begin{equation}
  m_0(\beta)=\sqrt{1-\frac{16 \exp(-12 \beta)}{[1-\exp(-4\beta)][1+3 \exp(-4
\beta)]}} \;.
\end{equation}
In contrast to the large number of low-temperature series expansions for the
square lattice, we are aware of only two published papers for the triangular
lattice \cite{sykes:73a,sykes:75}. In the second paper two more coefficients
for the same series are given:
\begin{equation}
  \chi(\beta) =  \beta\sum_{i=1}^n c_i u^{i}
  \hspace{0.5cm}\text{with}\hspace{0.5cm}
  u = \exp(-4 \beta) \;,
\end{equation}
where $c=\{$ 0, 0, 4, 0, 48, 16, 516, 288, 5\,328, 3\,840, 53\,676, 45\,488,
531\,600, 505\,584, 5\,199\,404, 5\,399\,136, 50\,369\,760, 56\,095\,776,
484\,296\,732, 571\,273\,344, 4\,628\,107\,216 $\}$. Finally, for the volume of
the Wulff plot no explicit solution is available. Shneidman and Zia
\cite{shneidman:01} showed the correct solution to be the integral
\begin{equation}
  W(\beta) = 6 \int\limits_0^{\pi/6} \mathrm d\theta \; r^2(\theta)
\end{equation}
with a function $r(\theta)$ given implicitly by
\begin{multline}
 \frac{3+\exp(2 \beta)}{-2+2 \exp(2\beta)} = 
      \cosh\left[r \beta \sin\left(\frac{\pi}{3}-\theta\right)\right]
      \\
      + \cosh\left[r \beta \sin\left(\theta\right)\right]
      + \cosh\left[r \beta \sin\left(\frac{\pi}{3}+\theta\right)\right] .
 \label{eq:r.indirect}
\end{multline}
For the angles $\theta_l=l\pi/6$, $l=0,1,\dots,11$ the interface tension in
direction normal to the equilibrium surface is given by $r(\theta_l)$.
In the direction $\theta=\pi/6$ the minimal radius $r_\text{min}$ can be
found to have the value
\begin{multline}
 r_\text{min} = \sigma_0 =  \\
 \frac 2 \beta \cosh^{-1} \left(
 \frac{1-e^{4\beta}+e^{2\beta}\sqrt{e^{8\beta}-2e^{4\beta}-3} }{2 e^{4\beta}-2}
 \right) \;.
 \label{eq:r.min}
\end{multline}
The maximal radius $r_\text{max}$ is located at $\theta=0$ and
Eq.~(\ref{eq:r.indirect}) simplifies greatly to 
\begin{equation}
 r_\text{max} = \sigma_1 = \frac{2}{\sqrt 3 \beta} \ln\left( \frac{e^{4
\beta}-1}2 \right) \;.
 \label{eq:r.max}
\end{equation}

\subsection{\label{sec:parameters.nnn}Parameters for the n.n.n.\ Ising model on
a square lattice}

\begin{figure}
 (a) \includegraphics[scale=0.7]{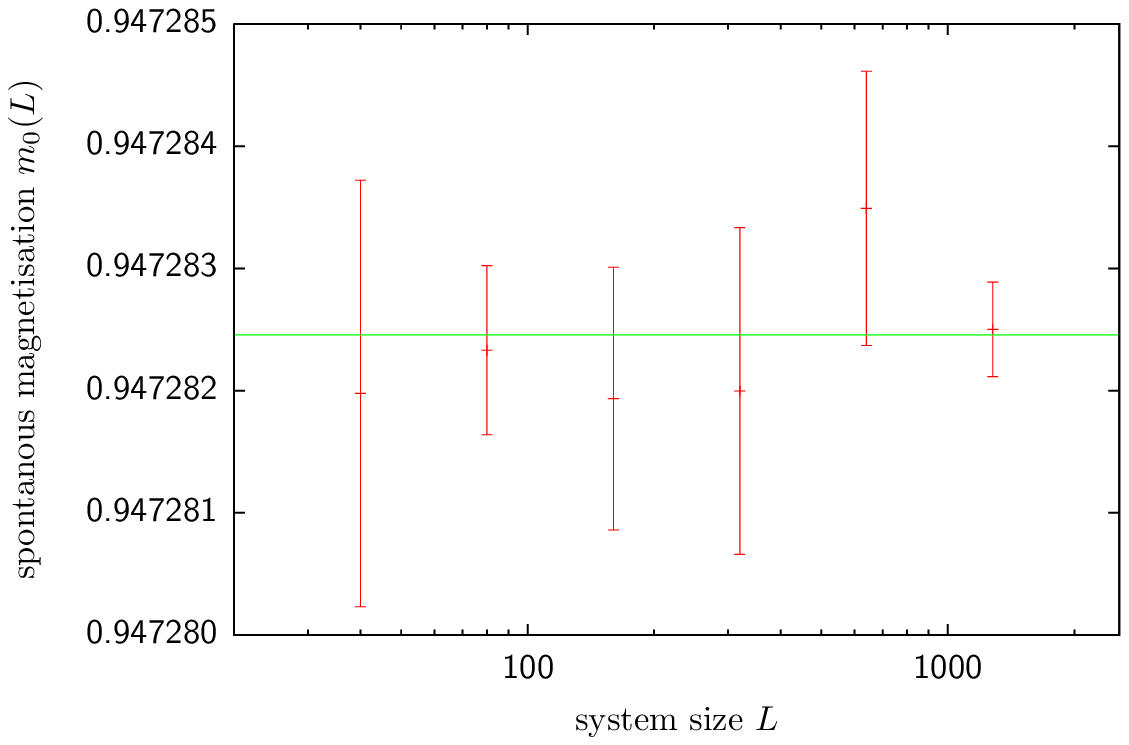}\\
 (b) \includegraphics[scale=0.7]{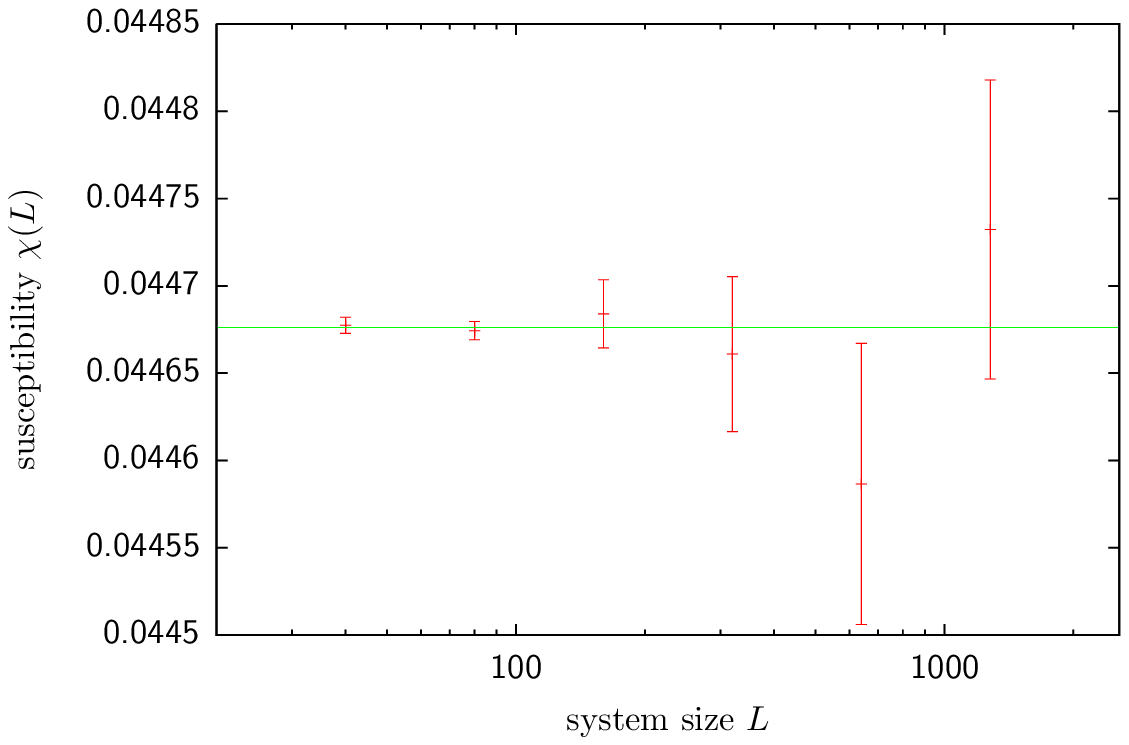}
 \caption{The horizontal (green) line marks the mean of (a) the spontaneous
magnetisation $m_0(L)$ and (b) the magnetic susceptibility $\chi(L)$ for system
sizes $L=40 \dots 1280$ at $T=4.0$ of a n.n.n.\ Ising model. Its value gives an
estimate for $m_0$ and $\chi$ at $L \to \infty$. Here, we read of the
values $m_0=0.947\,2825(2)$ and $\chi=0.044\,676(2)$.}
 \label{fig:fit.nnn.m0.chi}
\end{figure}

For the next-nearest neighbour model none of our parameters are known
exactly. The inverse critical temperature was given by Nightingale and Bl{\"o}te
\cite{nightingale.betac} using a transfer-matrix technique they call
``phenomenological renormalisation'' to be
\begin{equation}
 \beta_\mathrm c = 0.190\, 192\, 69 (5) \;.
\end{equation}
In \cite{nussbaumer:07} this value was independently established using Monte
Carlo simulations and finite-size scaling procedures.
All other quantities are unknown in the literature and, therefore, computer
simulations must provide the values. In the case of the magnetisation and the
magnetic susceptibility this is quite easy. A simple Monte Carlo algorithm at
the desired temperature gives a time series of the magnetisation $M$. Then, the
spontaneous magnetisation and the susceptibility are given by
\begin{equation}
  \label{eq:nnn.m0.chi}
  m_0 = \frac 1{VN} \sum_{i=1}^N M_i
\end{equation}
and
\begin{equation}
 \chi =\frac{\beta}{V} \left[\frac 1 N\sum_{i=1}^N M_i^2-\left(
\frac 1 N\sum_{i=1}^N
M_i\right)^2
\right] \;,
\end{equation}
where $N$ is the number of Monte Carlo measurements and $V=L \times L$ the
volume of the system. In the desired temperature range $T \approx (2/3)
T_\mathrm c$ the spatial correlation length $\xi$ is very
small and therefore already for moderate lattice sizes rather precise
estimates can be achieved \footnote{For too small lattice sizes, the system can
``tunnel'' from one peak of the magnetisation, e.g.\ at $m=m_0$ to the peak of
opposite magnetisation $m=-m_0$ or vise versa. To be on the safe side, we
checked the time series for this behavior.}. Figure~\ref{fig:fit.nnn.m0.chi}
shows the results of a Metropolis simulation of the n.n.n.\ Ising model at
$T=4.0$.

To obtain the Wulff free energy is a much more demanding task. Several methods
are known, e.g.\ thermodynamic integration
\cite{buerkner.stauffer,hasenbusch.pinn}. Here, we will discuss two different
ideas, namely a fit to the distribution of $P(M)$ and a simple argument that the
value of $\tau_\mathrm W$ does not differ much from the appropriately
scaled planar surface tension $\sigma_0$.

\begin{figure}
 (a) \includegraphics[scale=0.68]{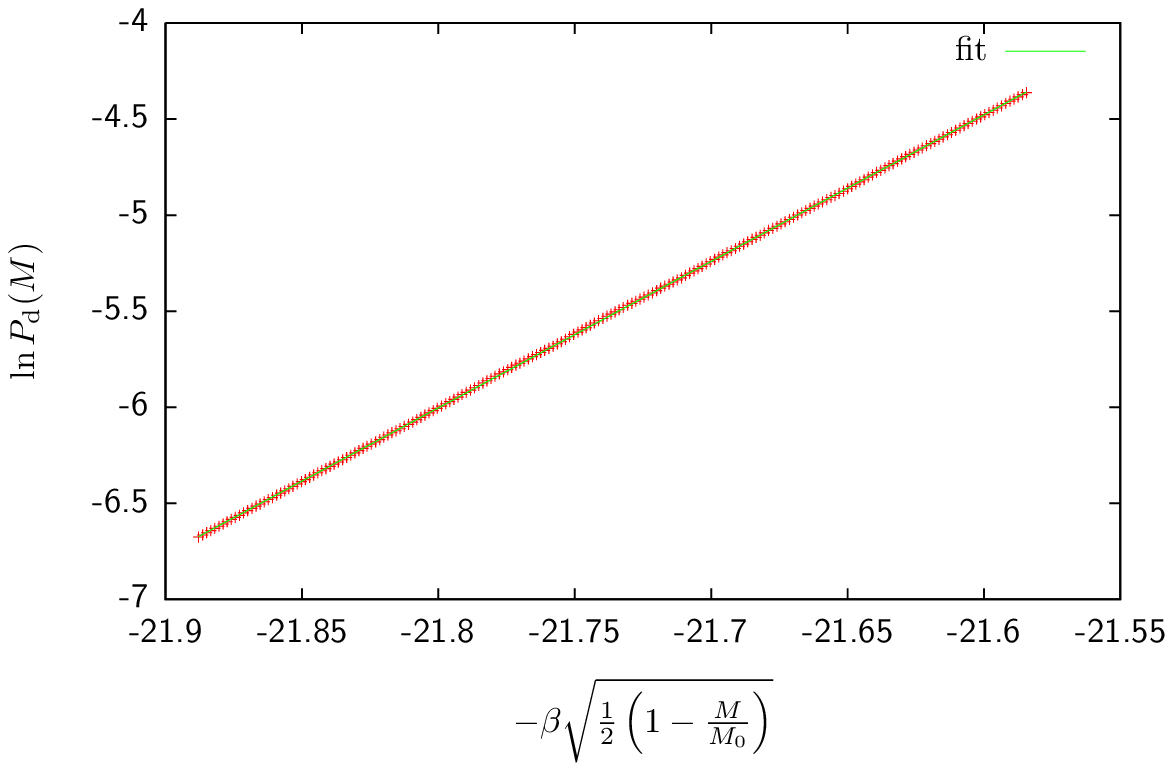}\\
 (b) \includegraphics[scale=0.68]{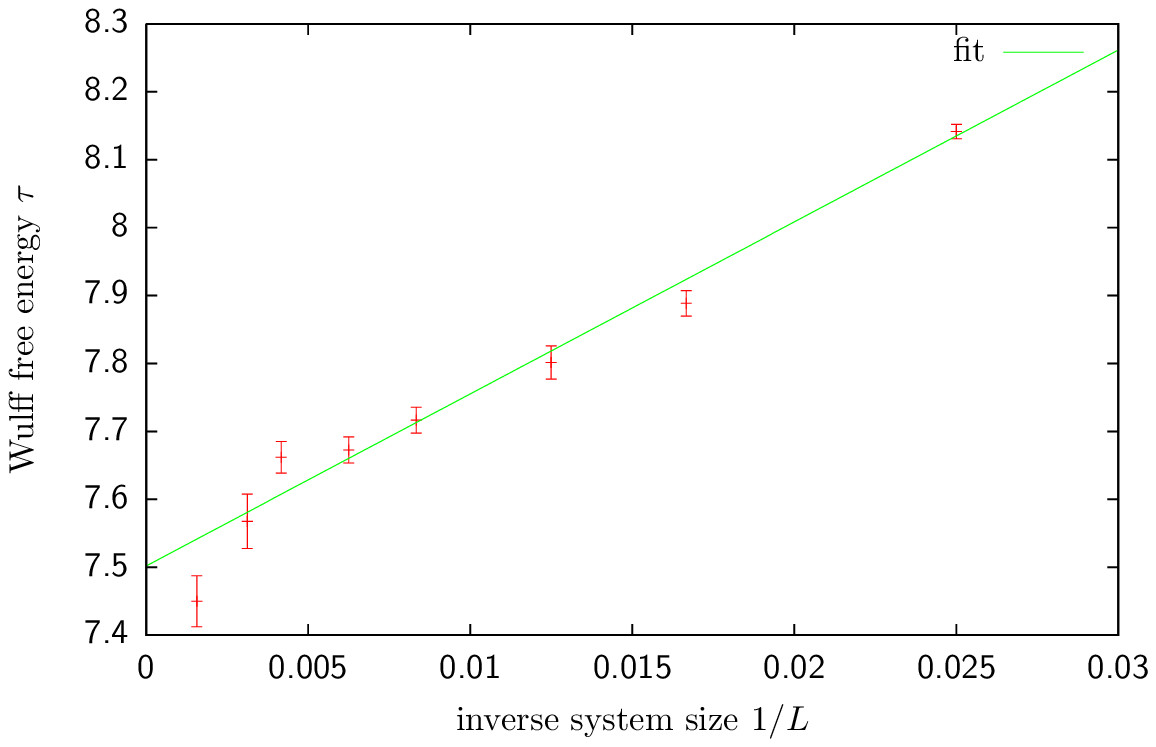}
 \caption{(a) Fit of the distribution $\ln P_\mathrm d(M) = -\beta
\tau_\mathrm W \sqrt{1/2 (1-M/M_0)}$ for a $V=160\times 160$ n.n.n.\ Ising model
at the temperature $T=4.0$ in the range $m=[0.4,0.4+400/160^2]$. (b) Fit of the
Wulff free energy $\tau_W$ vs. the inverse system size $L$ at temperature
$T=4.0$ for $L=40, 80, \dots, 640$. The error bars are obtained from (at
least) 10 independent simulations per data point.}
\label{fig:fit.log.PM}
\end{figure}

For our first method we exploit the fact that the probability distribution
for the largest droplet can be written as \cite{shlosman.tube}
\begin{equation}
  P_\mathrm d \propto \exp \left( -\beta \tau_\mathrm{W} \sqrt{v_\mathrm d} 
\right) \;.
  \label{eq:shlosman.droplet}
\end{equation}
Using Eq.~(\ref{eq:M.vL}) and under the assumption $v_\text d \approx v_L$ the
free energy in the exponent is
\begin{equation}
 F_\mathrm d
  = \tau_\text W \sqrt{v_\text d}
  \approx \tau_\text W \sqrt{\frac 12 \left( 1-\frac M {M_0} \right ) }
\;.
  \label{eq:f.shlosman.droplet}
\end{equation}
The assumption that the total overturned volume $v_L$ is consumed by the droplet
volume $v_\mathrm d$ is certainly fulfilled the better the larger the droplet
is. As is well known, the droplet can grow until it reaches the so-called
droplet/strip transition point which is roughly located at
\begin{equation}
  M_\mathrm{ds} = M_0 \left(1- \frac 2 \pi \right) \;.
  \label{eq:ds}
\end{equation}
With Eqs.~(\ref{eq:shlosman.droplet}) and (\ref{eq:f.shlosman.droplet}), a
linear fit of the form $y=\tau_\mathrm W x +c $ can be achieved, where $y=\log
P_\mathrm d$ and $x=-\beta \sqrt{1/2 (1-M/M_0)}$. Figure~\ref{fig:fit.log.PM}
(a) shows such a fit for the $160 \times 160$ n.n.n.\ Ising model at the
temperature $T=4.0$ and for a range $m=[0.4000,0.4156]$ which is close to the
droplet/strip transition point located at
$m_\text{ds}=m_0\left(1-2/\pi\right)\approx 0.3442$.
The data stems from a constrained multimagnetic simulation. To extract the value
of the Wulff free energy in the thermodynamic limit of large systems, several
simulations at different lattice sizes must be performed. In
Fig.~\ref{fig:fit.log.PM} (b) the scaling of the Wulff free energy is
shown in dependence of the inverse lattice size. The intersection of the linear
fit with the ordinate gives an estimate of $\tau_\mathrm W = 7.50 \pm 0.02$.

Finally, we want to make three remarks about the given method. Firstly, we are
fully aware of the fact, that Eqs.~(\ref{eq:lambda.nontrivial}) and
(\ref{eq:lambda}) give a ``correction'' to the fit done last. Using
$v_\text d(\lambda)$ the fit would be valid for any droplet size up to the
condensation/evaporation point and not only for large droplets nearby the
droplet/strip transition point. But on the other hand, the fit would not be a
linear anymore and more important, the theoretical predictions that we want to
compare with would mix up with the parameter estimation. Secondly, it is
possible to measure during the simulation the  droplet size $v_\text d$ and fit
directly $\tau_\mathrm W \sqrt v_\mathrm d$ instead of $P(m)$. Here, the
disadvantage lies in the computational effort to measure the droplet size. While
the magnetisation comes at no additional cost, a single measurement of the
volume of the largest droplet needs ${\cal O}(V)$ operations. Thirdly, we want
to emphasise the importance of the initial starting conditions of the
simulation. An ordered start where the first $n$ spins point in one direction
and the next $V-n$ in the other direction is in fact a strip configuration. As
discussed in \cite{leung:90,neuhaus.hager} between the strip configuration and
the droplet configuration there is an exponentially large barrier that might not
be overcome during the equilibration phase, even so a droplet configuration has
a much lower free energy for the constrained magnetisation range chosen.

\begin{figure}

 (a) \includegraphics[scale=0.7]{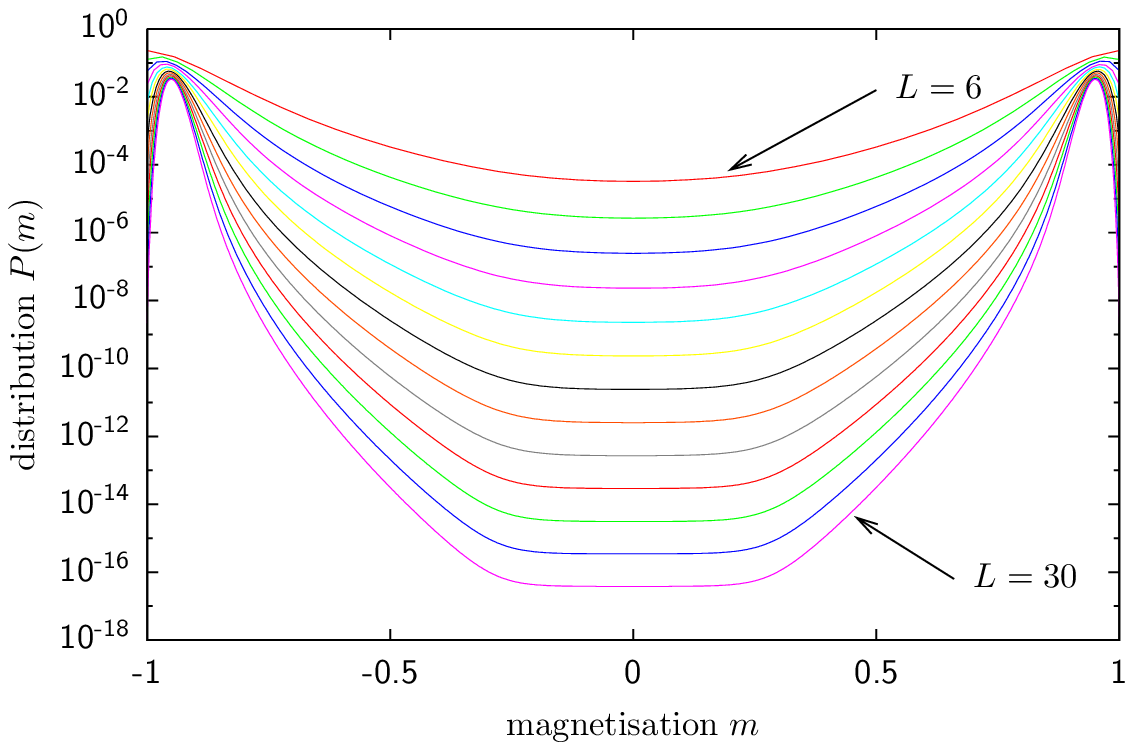}\\
 (b) \includegraphics[scale=0.7]{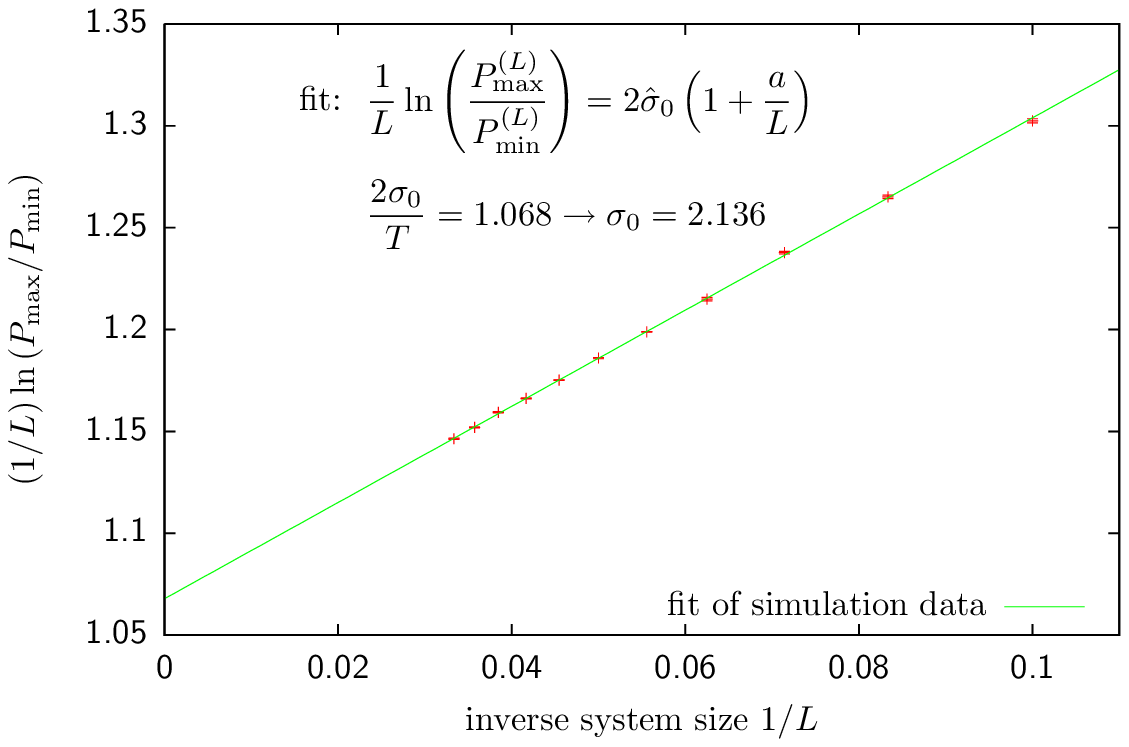}
 \caption{
 (a) Distribution of the magnetisation $m$ for the n.n.n.\ Ising model at
$T=4.0$ and system sizes $L=6,8, \dots, 30$.
 (b) Scaling of the interface-tension estimates from the histogram method: The
straight line shows the fit $\ln(P_\mathrm{max}^{(L)}/P_\mathrm{min}^{(L)})/L=2
\beta \sigma_0(1+a/L)$  for $L \ge 10$ with goodness-of-fit parameter
$\chi^2/\text{d.o.f.}=1.1$, yielding an planar interface tension estimate of
$\sigma_0=2.136 \pm 0.001$. }
 \label{fig:fit.mum}
\end{figure}

The second method to obtain $\tau_\mathrm W$ is based on the assumption that, at
the considered temperature, the interface tension for different angles $\theta$
is roughly isotropic. This can be verified in detail for the n.n.\ Ising model,
where the interface tension for an arbitrary angle $\theta$ is known
analytically \cite{avron.sigma0}. For the planar interface the expression (also
given by Onsager \cite{onsager:44,mccoy.wu,baxter:82}) is $\sigma_0^\text{sq}=2
J+T \ln \left[\tanh(J/T)\right]$ and the expression for the ``worst case'',
i.e.\ along the main diagonal of the lattice, is $\sigma_1^\text{sq}=\sqrt 2 T
\ln \sinh(2 J/T)$ (also given by Fisher and Ferdinand \cite{fisher:67}). For all
temperatures larger than $T=1.5$, the relative difference of
$\sigma_0^\text{sq}$ and $\sigma_1^\text{sq}$
is smaller than $1.3\%$. Obviously, the
Wulff shape is still rather circular at low temperatures and the quadratic form
becomes only apparent close to  $T=0$.  With this crude heuristics, the
interface tension per unit volume at $T=1.5$ is $2 \sqrt{\pi} \sigma_0^\text{sq}
= 4.219$. This is quite close ($99.37$\%) to the correct value
$\tau_{\text W}^\text{sq}=4.245$. An even better approximation is $2\sqrt \pi
(\sigma_0^\text{sq} + \sigma_1^\text{sq})/2$ that deviates only $0.006 \%$ from 
the actual value. The same holds true for the triangular lattice. Using
Eq.~(\ref{eq:r.min}) one finds at $T=2.4 \approx \frac 2 3 T_\text c$ a value of
$2 \sqrt \pi \sigma_0^\text{tri}=7.50657$ which is only $0.005 \%$ smaller than
the exact value of $\tau_\text W^\text{tri}$. Including Eq.~(\ref{eq:r.max}) for
the improved estimation $2\sqrt \pi (\sigma_0^\text{tri} +
\sigma_1^\text{tri})/2$ yields a remarkably small difference of tiny $6 \times
10^{-7} \%$ to the exact result. A more detailed discussion concerning the
approximation of $\sigma(\theta)$ can be found in \cite{shneidman:01}. For the
n.n.n.\ Ising droplet the low-temperature Wulff shape is an octagon, i.e.\ it is
much closer to the high temperature (low
interface tension) form, namely a circle. Therefore, it is reasonable to assume
that above approximation might work as well. The planar interface tension can be
measured using a multimagnetical (flat in the distribution of the magnetisation)
simulation, the result of which is a double-peaked magnetisation density $P(m)$.
In the limit of large system sizes $L$, it holds in two dimensions
\cite{janke.nato}
\begin{equation}
 \ln \left(\frac{P_\mathrm{max}^{(L)}}{P_\mathrm{min}^{(L)}} \right)
 = 2 \beta \sigma_0 L \;,
 \label{eq:max.min}
\end{equation}
where $P_\mathrm{min}^{(L)}$ is the value of the density in the mixed phase
region $m \approx 0$ and $P_\mathrm{max}^{(L)}$ the value at its maxima ($m =
\pm m_0$). Figure~\ref{fig:fit.mum} (a) shows the result of 13 multimagnetic
simulations for the systems sizes $L=6$ to $L=30$ \cite{nussbaumer:07}. For
every system the maximum and minimum probability $P_\mathrm{max}^{(L)}$ and
$P_\mathrm{min}^{(L)}$ were read off and repeating the simulations ten times
error bars were obtained. For $L \ge 10$ the resulting values are plotted in
Fig.~\ref{fig:fit.mum} (b). An infinite system size extrapolation in $1/L$
yields a value of 
$\sigma_0=2.136 \pm 0.001$ for the planar interface tension. Then, the estimate
for the Wulff free energy (assuming a circular droplet shape) is
$\tau_\mathrm{W} \approx 2 \sqrt{\pi} \times 2.136 = 7.571 \pm 0.004$ which in
fact is a lower bound, as the interface tension gets minimal along the
directions of the interactions.

Table~\ref{tab:numeric} gives the numerical values for the spontaneous
magnetisation $m_0$, the susceptibility $\chi$ and the Wulff free energy at the
temperature $T$ were the simulation took place. The temperature was chosen to
be $T \approx 0.66 \, T_\text c$ -- $0.76 \, T_\text c$ which is a good
compromise between simulation speed (freezing at low temperatures) and
compactness of the droplet (see the r.h.s.\ of Fig.~\ref{fig:ce} for a typical
configuration).

\begin{table}
 \caption{Numerical values for the magnetisation $m_0$, susceptibility $\chi$
 and Wulff interfacial free energy density $\tau_\mathrm{W}$ entering
 the parameters $\Delta = \Delta(v_L, m_0, \chi, \tau_\mathrm{W})$ defined in
 Eqs.~(\ref{eq:nn.m0}) to (\ref{eq:nnn.m0.chi}) at the simulation temperature
 $T$ for the three models studied.}
 \label{tab:numeric}
 \begin{tabular}{lddd}
  \hline\hline
                 & \text{n.n.\ sq.} & \text{n.n.\ tri.} & \text{n.n.n.\ sq.} \\
  \hline
  $T_\text c$                & 2.269   & 3.641   & 5.258   \\
  $T$                        & 1.500   & 2.400   & 4.000   \\
  $T/T_\text c$              & 0.6610  & 0.6592  & 0.7608\footnote{The
temperature $T=4.0$ was chosen without the knowledge of the critical
temperature, certainly a value of $T=3.5$ would have been more appropriate.}\\
  $m_0$                      & 0.9865  & 0.9829  & 0.9473  \\
  $\chi$                     & 0.02708 & 0.01959 & 0.04467 \\
  $\tau_\text W$             & 4.245   & 7.507   & 7.502   \\
  $2m_0^2/\tau_\text W \chi$ & 16.93   & 13.14   & 5.307 \\
  \hline\hline
 \end{tabular}
\end{table}

\subsection{Correction of the units in the parameter $\Delta$}
\label{sec:correction.of.units}

After all constants are known, there are still some considerations to be made,
before the parameter $\Delta$ can be calculated. The magnetisation $m_0$ and the
susceptibility $\chi$ are intensive quantities that follow from the
corresponding extensive quantities normalised (divided) by the volume. It is
convention that for spin systems the volume is expressed by the number of
spins, i.e.\ every spins accounts for a unit volume. In contrast, the free
energy of the Wulff droplet is measured (again by convention) in units of
the cell volume that is calculated given the lattice spacing $a$ as
input. As possible way to treat this situation is to normalise all quantities
to cell volume, which would mean, that $m_0$ and $\chi$ are given in very
unfamiliar units. We refrain from this step in order to keep things comparable
to literature and instead modify Eq.~(\ref{eq:Delta}) in a very slight way.
In order to do so, we define a scaling parameter $\Delta_\mathrm{lit}$ where
all parameters are consistent with the conventions from literature
\begin{equation}
 \Delta_\mathrm{lit} = 2 \frac{m_0^2}{\chi \tau_\mathrm W}
\frac{v_L^{3/2}}{L^2} \;.
 \label{eq:Delta.lit}
\end{equation}
Here, $v_L$ is the {\it number} of spins of the largest droplet including
overturned spins, $L^2$ is the {\it total number} of spins of the system and
$m_0$, $\chi$ are the magnetisation and susceptibility normalised to the {\it
total number of spins}. The normalisation of the Wulff free energy $\tau_\mathrm
W$ does not change as it is given in terms of the unit volume in literature.
Secondly we define $\Delta_\mathrm{uv}$ where all quantities are given in terms
of the unit volume which is the intended meaning by Biskup \textit{et al.},
\begin{equation}
 \Delta_\mathrm{uv} = 2 \frac{\mu_0^2}{X \tau_\mathrm W} \frac{\Omega^{3/2}}{V}
\;.
 \label{eq:Delta.uv}
\end{equation}
In this representation $\Omega$ is the {\it volume} of the largest droplet, $V$
the {\it volume} of the total system, and $\mu_0$ and $X$ are the magnetisation
and susceptibility normalised to the {\it volume} of the total system. If $v_0$
is the Voronoi volume of one spin \cite{voronoi} (the volume of the Wigner-Seitz
cell of one spin) measured in units compatible with $\tau_\mathrm W$, then it
holds
\begin{align}
 \Omega & = v_L v_0 \;, \\
 V      & = L^2 v_0 \;, \\
 \mu_0  & = \frac{M}{V} = \frac{M}{v_0 L^2} = \frac{m_0}{v_0} \;, \\
 X      & = \beta V \left( \left< \mu^2 \right>-\left< \mu\right>^2 \right) \\
        & = \beta L^2 v_0 \left( \left<\frac{m_0^2}{v_0^2} \right>-
          \left< \frac{m_0}{v_0} \right>^ 2\right) \\
        & = \frac{\beta L^2}{v_0} \left( \left< m_0^2 \right> -
          \left< m_0\right>^2 \right) = \frac{\chi}{v_0} \;.
 \label{eq:Delta.trans}
\end{align}
Now, a geometric ``correction factor'' $\alpha$ from $\Delta_\mathrm{lit}$ to
$\Delta_\mathrm{uv}$ can be defined as
\begin{equation}
 \Delta_\mathrm{uv} = \alpha \Delta_\mathrm{lit} \;.
 \label{eq:Delta.alpha} 
\end{equation}
Using Eqs.~(\ref{eq:Delta.lit})-(\ref{eq:Delta.alpha}) $\alpha$ can be expressed
as
\begin{align}
 \alpha
 = \frac{\Delta_\mathrm{uv}}{\Delta_\mathrm{lit}}
 = \frac{2 \frac{\mu_0^2}{X \tau_\mathrm W} \frac{\Omega^{3/2}}{V}}{2
\frac{m_0^2}{\chi \tau_\mathrm W} \frac{v_L^{3/2}}{L^2} }
 = \frac{\left(\frac{m_0^2}{v_0}\right) (v_L v_0)^{3/2} L^2}{\frac{X}{v_0}
 m_0^2 L^2 v_0 v_L^{3/2}} = \frac{1}{\sqrt{v_0}} \;.
\end{align}
To conclude, using the parameters from literature as given in
Table~\ref{tab:numeric}, the abscissa must not be scaled with $\Delta$ but
rather with $\Delta / \sqrt{v_0}$ where $v_0$ is the Voronoi volume of one cell.

\begin{figure}
  (a) \includegraphics{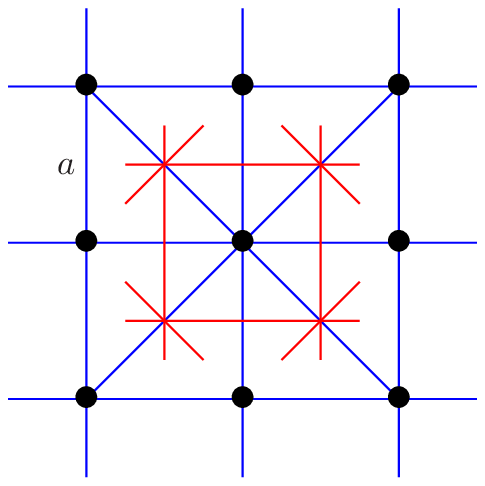}\\
  (b) \includegraphics{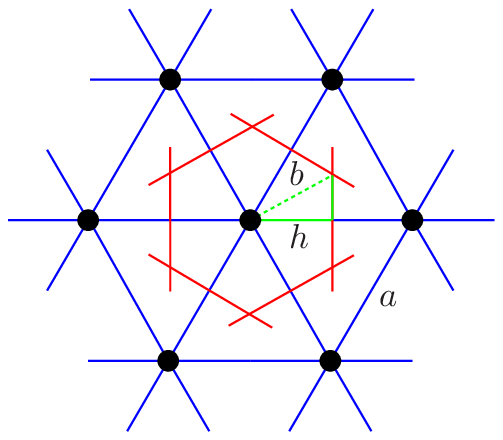}
  \caption{The Wigner-Seitz cell of the (a) n.n.n.\  and (b) triangular
lattice. It contains only one lattice site and all points within the cell are
closer to this point than to any other lattice site. The red lines indicate the
construction principle using the normals to the connection of a lattice to
its neighbours.}
  \label{fig:wigner.seitz.cell}
\end{figure}

For the square lattice the Voronoi volume that a spin occupies is $1
\times 1$ which makes the correction factor transparent. The same holds for the
n.n.n.\ lattice that has (by incident) the same geometry as the n.n.\ lattice,
see Fig.~\ref{fig:wigner.seitz.cell}.
In the case of the triangular lattice the Voronoi cell is a hexagon.
Figure~\ref{fig:wigner.seitz.cell} (b) displays the situation. If $h$
denotes the half of the lattice side length $a$, then it holds $a=2h$. Every
hexagon is made up of 6 small equilateral triangles of side length $b$ (dotted
line). The height of such a triangle is $h$ which is given by $h=b\sqrt 3/2$. It
follows, that $b=a/\sqrt 3$. Now, the volume of a hexagon is given by
\begin{equation}
 v_0^\mathrm{hex} =  \frac{3 \sqrt 3} 2 b^2
 = \frac{3 \sqrt 3}{2} \left(\frac{a}{ \sqrt 3} \right)^2 = \frac{\sqrt
3}{2}a^2 \;.
\end{equation}
Finally, for $a=1$, the geometric factor $\alpha$ for the triangular lattice is
\begin{equation}
 \alpha_\text{tri}= \frac{1}{\sqrt{v_0}} = \sqrt \frac{2}{\sqrt 3} \approx
1.075 \dots \;.
\end{equation}


\subsection{\label{sec:methods}Droplet measurement}

As mentioned at the beginning of Sec.~\ref{sec:setup}, one of our primary
goals was the determination of the volume of the largest droplet.
A possible advancement would be to set up a multimagnetic simulation and measure
every sweep or so the droplet volume. While this is certainly possible, it is
not advisable, as the determination of the multimagnetic weight factors $W(m)
\approx 1/P(m)$ alone is a demanding task and in the following analysis there is
no use for them. Instead, we arranged several simulations at fixed magnetisation
$m$ (micromagnetic). Inserting Eq.~(\ref{eq:deltaM.vL}) in (\ref{eq:Delta}) and
solving for $M$ gives the relation between the parameter $\Delta$ and the
magnetisation $M$
\begin{equation}
  M(\Delta) = V m_0 -
    \left(\frac{2 \Delta \chi \tau_\mathrm{W} V}{\sqrt{2 m_0}} \right)^{2/3} \;.
  \label{eq:M.Delta}
\end{equation}
Solving Eq.~(\ref{eq:M.Delta}) for $\Delta$ yields
\begin{equation}
 \Delta(M) = \frac{\sqrt{2 m_0}}{2 \chi \tau_\text W V} \left(V m_0 -
M\right)^{3/2} \;,
 \label{eq:Delta.M}
\end{equation}
which shows that a fixed magnetisation results in a fixed value
$\Delta(M)$. Therefore, we actually selected for every lattice 38 reasonable
values $\tilde \Delta_i = \left\{0.00, 0.10, \dots 16\right\}$, with an emphasis
on the vicinity of $\Delta_\mathrm{c}$. Using Eq.~(\ref{eq:Delta.M}) a set of
corresponding magnetisation values $M_i$, usually non-integer values, was
obtained. A subsequent rounding to the next allowed value of the magnetisation
($\Delta M = \pm 2$) gave the final values for the simulation. To take the
influence of the rounding into account, Eq.~(\ref{eq:M.Delta}) was used,
resulting in a second set $\Delta$ of slighty shifted ($\propto 1/\sqrt
V$) values $\Delta_i$ that correspond to the rounded magnetisation.

To enforce the constraint of constant magnetisation we use a Kawasaki update
scheme where an up-spin is exchanged with a down-spin. Since the total number of
up- and down-spins does not change, the magnetisation keeps its value as well.
This type of non-local Monte Carlo moves can be accelerated using a table
storing the spins sorted according to their direction. Here, one sweep accounts
for $V$ spin exchange attempts.

\begin{figure}
 (a) \includegraphics{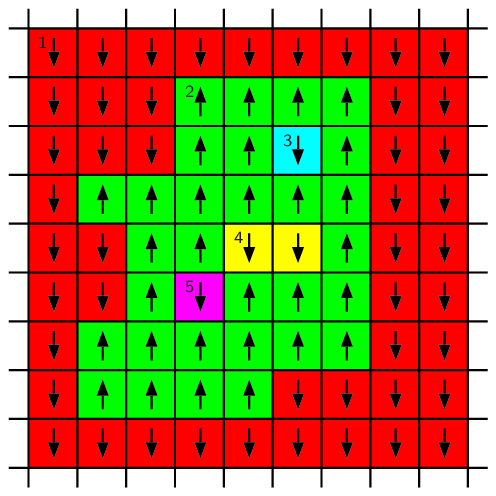}\\
 (b) \includegraphics{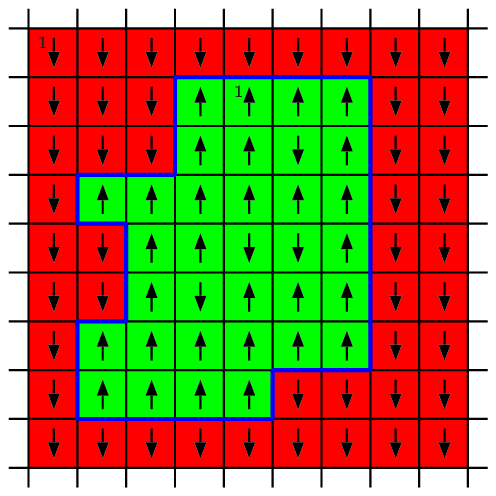}
 \caption{Cut out of a spin field, the red background cluster should be much
larger, cf.\ the r.h.s. of Fig.~\ref{fig:ce}. (a) The colors and the
(small) numbers indicate the clusters detected and enumerated by the
Hoshen-Kopelman routine. (b) The thick blue line surrounds the droplet
(second largest cluster) found by the flood-fill routine.}
 \label{fig:simple.fill}
\end{figure}

\begin{figure}
 (a) \includegraphics{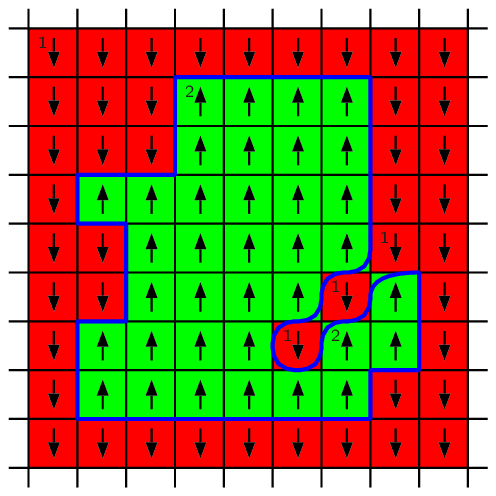}\\
 (b) \includegraphics{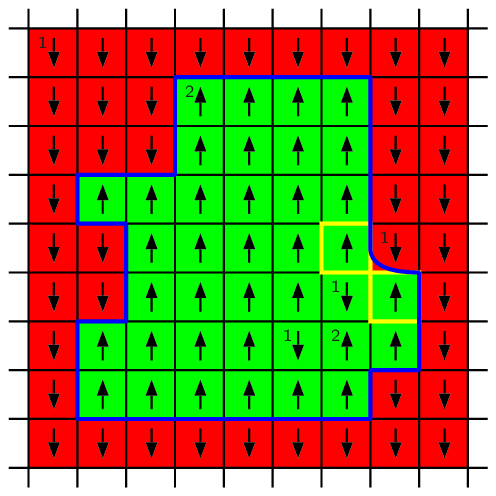}
 \caption{(a) Cut out of a spin field for the n.n.n.\ Ising model with a
droplet (green) detected by the flood-fill algorithm. Apparently, the inclusion
on the lower right side of the droplet (two spins) has a connection to the
backgrund and does not count to the volume of the droplet. (b) Another way
to interpret the situation where the spins are part of the droplet.}
 \label{fig:nnn.ff.problem}
\end{figure}

After every sweep our simulation determines the volume of the second-largest
cluster which is (per definition) the volume $v_\text d$ of the droplet. This is
done in two steps. First a Hoshen-Kopelman \cite{hoshen.kopelman} algorithm
performs a complete cluster decomposition. Thereby spins that are connected in
the sense that they share a bond and have like orientation become a unique
number. Figure~\ref{fig:simple.fill} shows the situation for a spin-field and
n.n.\ interaction. The largest (partially drawn) cluster (red) having cluster
index $1$ is the background, the cluster in the center (green) with cluster
index 2 is the droplet we are looking for. Inside this droplet are smaller
clusters located with cluster index 3, 4 and 5 (light blue, yellow, purple). In
the next step a flood-fill routine \cite{floodfill}, essentially a geometric
depth first search, scans the droplet. Starting from an arbitrary position (that
was recorded during the cluster identification step) it stops only when it finds
spins that belong to the largest cluster (background). Thereby spins/clusters of
opposite sign that lie within the droplet are subsumed. The result of this
operation is shown in Fig.~\ref{fig:simple.fill} (b). The thick blue line
indicates the border between the droplet, i.e., cluster number 2 and all
clusters which do not have the cluster number of the background, and the
background. While this method is easy to implement and for the n.n.\ square
lattice fool-proof, in case of the n.n.n.\ square lattice there are some
pathological cases. Figure~\ref{fig:nnn.ff.problem} shows such an ambiguous
situation. Figure~\ref{fig:nnn.ff.problem} (a) presents the droplet as
identified by our algorithm. In contrast, Fig.~\ref{fig:nnn.ff.problem} (b) is
an (imaginary) alternative version resulting from the closing of the inclusion
of background spins. The justification of the right pictures is given by the
fact that the n.n.n.\ model has an interaction along the diagonal which connects
the two surface spins (yellow). Fortunately, it is not necessary to decide upon
which scenario is the more physical one. Every inclusion of reasonable size
causes a large number of broken bonds due its surface. Therefore, configurations
with inclusions are highly suppressed for temperatures well below the Curie
point. To be on the safe side, we analysed several simulations of the n.n.n.\
square lattice for different system sizes with both methods at the same time,
i.e., for identical configurations the droplet was measured a second time with
an algorithm that closes inclusions, to find negligibe differences. In the end
we deciced to keep things as simple as possible and therefore used only the
combination Hoshen-Kopelman/flood-fill for our data generation.

\section{\label{sec:numerical.results}Numerical results}

For all three systems and at every value of $\Delta$ we performed simulations at
five different lattices sizes $L=40, 80, 160, 320$, and $640$. Every
simulation ran at least $20\, 000$ sweeps for the thermalisation and at
least $200\, 000$ sweeps for the measurements. To obtain the error bars
reliably, 10 independent simulations were run for each data point. For the
creation of pseudo random numbers we use the R250/521 generator
\cite{heuer:97,janke.rnd}.

\begin{figure}
  \includegraphics[scale=0.7]{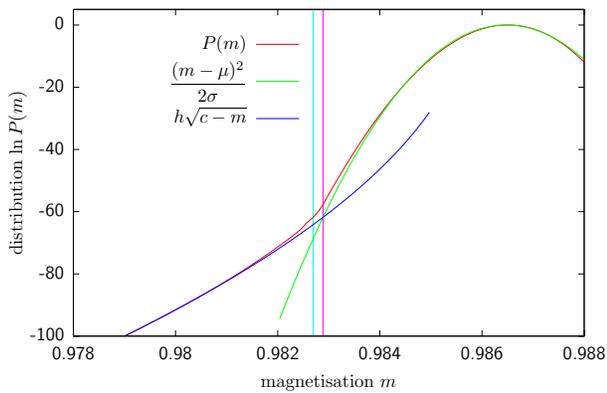}
  \caption{Gaussian fit and stretched exponential fit of the distribution of the
magnetisation $P(m)$ for a the result of a $L=640$ n.n.\ Ising simulation at
$T=1.5$. The left vertical line (magenta) indicates the transition
magnetisation $M(\Delta_\text c)/V$ predicted by Eq.~(\ref{eq:M.Delta}) while
the the right vertical line (purple) coincides with the intersection point of
the two fits.}
  \label{fig:fit.P_M}
\end{figure}

Having the numerical values of $m_0$, $\chi$, and $\tau_\text W$ in place (see
Sec.~\ref{sec:setup}), the region of interest can be estimated. For $\Delta=0.92
\approx \Delta_c$ and the values from Table~\ref{tab:numeric} corresponding to
the n.n.\ Ising model, for $L=640$ the magnetisation is estimated with
Eq.~(\ref{eq:M.Delta}) to be $m \approx 0.9827$. To see the
relevance of this figure we performed a multimagnetic simulation coupled with
the parallel tempering
algorithm \cite{earl.deem} for the n.n.\ Ising model, the result of which can
be seen Fig.~\ref{fig:cusp}. It shows the upper part (in the
vicinity of the magnetisation peak in Fig.~\ref{fig:cusp.schematic}) of the
distribution of the magnetisation $P(m)$ that exhibits for larger
lattice sizes a clear cusp which divides the evaporated and condensed region.
Within the evaporated region it has a Gaussian form according to
Eq.~(\ref{eq:F.f}), while in the condensed region a stretched
exponential behavior is visible, cf.\ Eq.~(\ref{eq:F.d}). To verify this
quantitatively, Fig.~\ref{fig:fit.P_M} shows a fit of a Gaussian curve and a
stretched exponential curve to the upper part of the distribution of the
magnetisation $\ln P(m)$ for the n.n.\ Ising model. The point of intersection
$m_\times$ is given by the condition
\begin{equation}
  h \sqrt{c - m_\times}+d = -\frac{(m_\times-m_\text{max})^2}{2 \sigma^2}
\end{equation}
the solution of which is a fourth order equation. With the parameters from the
fit $m_\text{max}=0.9864$, $\sigma^2=1.042 \times 10^{-7}$, $c=0.9858$ and
$h=-1340$, it evaluates to $m_\times=0.9829$, which is
quite close to the aforementioned value calculated with Eq.~(\ref{eq:M.Delta}).
The Gaussian fit which corresponds to the pure fluctuations part where
$\lambda=0$ can be compared to $- \beta F_\text f$ from Eq.~(\ref{eq:F.f}). It
yields for the susceptibility $\chi = \beta V \sigma^2=0.6666 \times 640^2
\times 1.042 \times 10^{-7} \approx 0.028$, a value quite close to the
infinite-volume value given in Table~\ref{tab:numeric} of $0.02708$. In the
droplet dominated regime we have approximated the full mixed phase expression by
neglecting the contributions of the fluctuations, which corresponds to putting
$\lambda=1$ in Eq.~(\ref{eq:F.total}). Over the fit range the neglected part
contributes less than $4 \%$. Even in the worst case, located at the cusp where
$\lambda=2/3$, it amounts only to a value of approximately  $9 \%$. To obtain
these values the ratio $F_\text d(1)/F(\lambda) = 4 \sqrt \lambda /(3 \lambda
+1)$ is evaluated using Eq.~(\ref{eq:lambda.nontrivial}) in conjunction with
Eq.~(\ref{eq:Delta.M}) which yields an expression $\lambda=\lambda(M)$. This
is corroborated by the fact that, when fitting the droplet
regime without fluctuations, from $-\beta F_\text d(1)$ the Wulff free energy
is approximated as $\tau_\text W = -h /[\beta \sqrt{V/(2 c)}]=1340/[0.6666
\times \sqrt{640^2/(2 \times 0.9858)}] \approx 4.410$, which is, again, quite
close to the value of $4.245$ given in Table~\ref{tab:numeric}.

To have another ``visual proof'' that something different is happening on the
two sides of the cusp in Fig.~\ref{fig:cusp} we took several snapshots of the
configurations that occurred during a simulation run. The two plots of
Fig.~\ref{fig:ce} display an evaporated (left) and a condensed system
(right), respectively.  Both systems have the same number of overturned spins,
i.e.\ the same magnetisation, which was chosen to be right at the transition
point. While both configurations occured during an actual simulation run,
they do present extreme cases. When looked at the set of the largest
cluster sizes recorded in the simulation run, the evaporated cluster 
configuration corresponds to the smallest number in the set and the condensed
configuration corresponds to the largest number in the set.

\begin{figure*}
  \includegraphics[scale=.68]{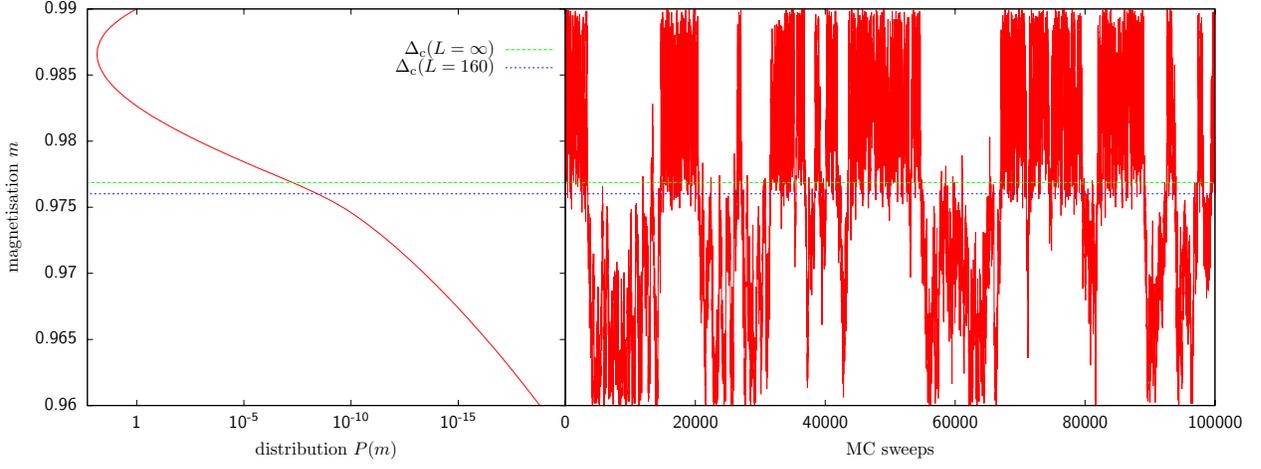}
  \caption{Time series of the magnetisation of a $L=160$ n.n.\ Ising
simulation at temperature $T=1.5$. The distribution on the l.h.s.\ corresponds
to the times series on the r.h.s\ and both were measured during the same
simulation run. The lower (green) and upper (blue) horizontal lines indicate the
transition magnetisation calculated from Eq.~(\ref{eq:M.Delta}) for a lattice
size $L=160$ and $L \to \infty$, respectively. The blocks in the time series
are typical sign of a barrier in the free energy.}
  \label{fig:fit.ts.P_M}
\end{figure*}

A final affirmation that the point under consideration was chosen
correctly, can be derived from a look at the time series of the magnetisation
$m$ in Fig.~\ref{fig:fit.ts.P_M}. The direct comparison shows a block structure
in the time series that coincides with cusp in the distribution $P(m)$.
Clearly, a sign for a barrier in the free energy.

\begin{figure}
 (a) \includegraphics[scale=0.7]{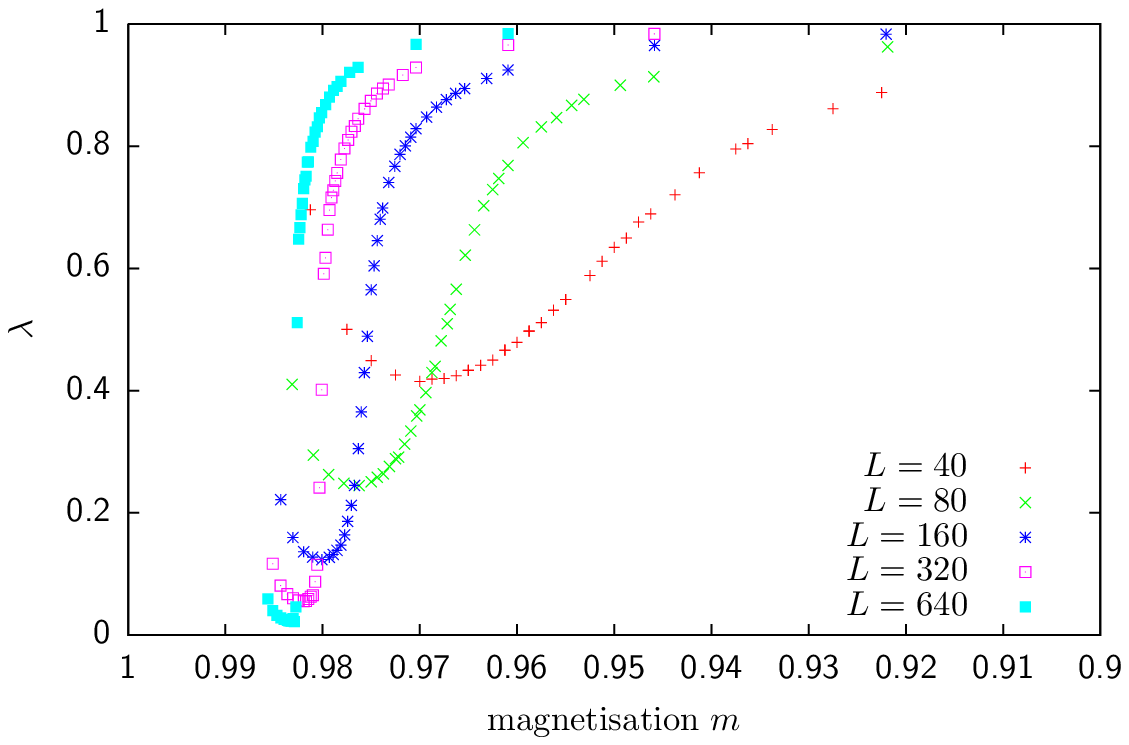}\\
 (b) \includegraphics[scale=0.7]{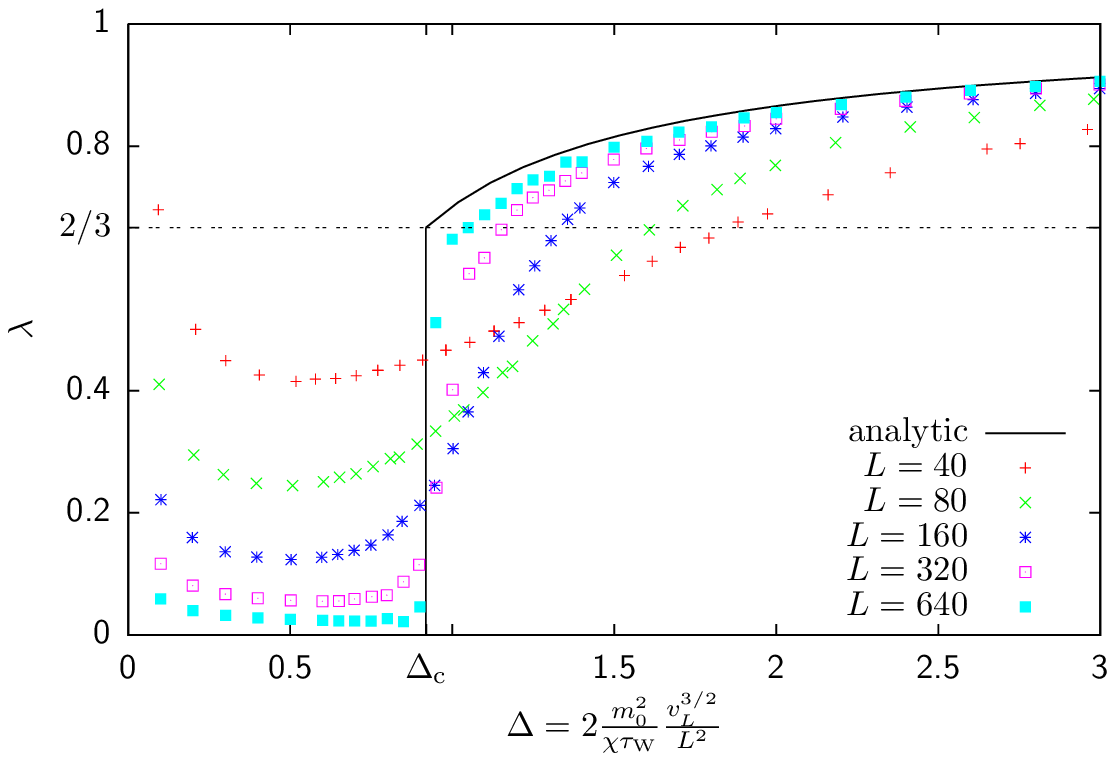}
 \caption{Fraction $\lambda$ for the two-dimensional n.n.\ Ising model on square
lattices of size $L=40, 80, \dots, 640$ with periodic boundary conditions at the
temperature $T=1.5 \approx 0.66 \, T_c$. The error bars are not plotted since
their size is much smaller than that of the data symbols. To show the influence
of the scaling of the absissa, plot (a) and (b) use the same date. While in
plot (a) the fraction $\lambda$ is given in units of the
magnetisation in plot (b) it is given in units of $\Delta$. The solid line
in plot (b) shows the analytic solution in the limit $L \to \infty$.}
\label{fig:nn.lambda.delta}
\end{figure}

\begin{figure}
  \includegraphics[scale=0.7]{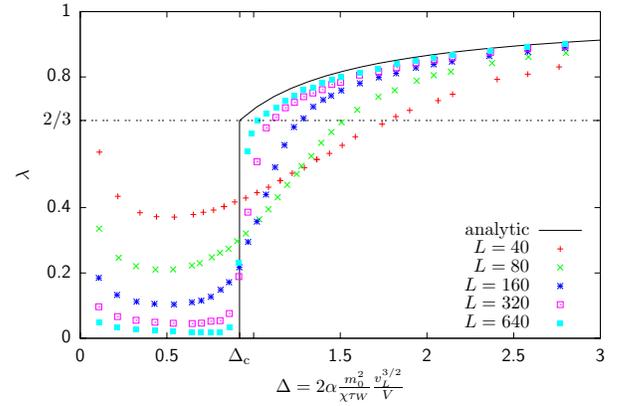}
  \caption{Fraction $\lambda$ for the two-dimensional triangular Ising model on
square lattices of size $L=40, 80, \dots, 640$ with periodic boundary conditions
at the temperature $T=2.4 \approx 0.66 \, T_c$. Here,
$\alpha=1/\sqrt{v_0}=\sqrt{2 /\sqrt{3}} \approx 1.075 \dots$ is the geometric
factor, defined
in Sec.~\ref{sec:correction.of.units}. The error bars are not plotted
since their size is much smaller than that of the data symbols. The solid line
shows the analytic solution in the limit $L \to \infty$.}
  \label{fig:tri.lambda.delta}
\end{figure}

\begin{figure}
  \includegraphics[scale=0.7]{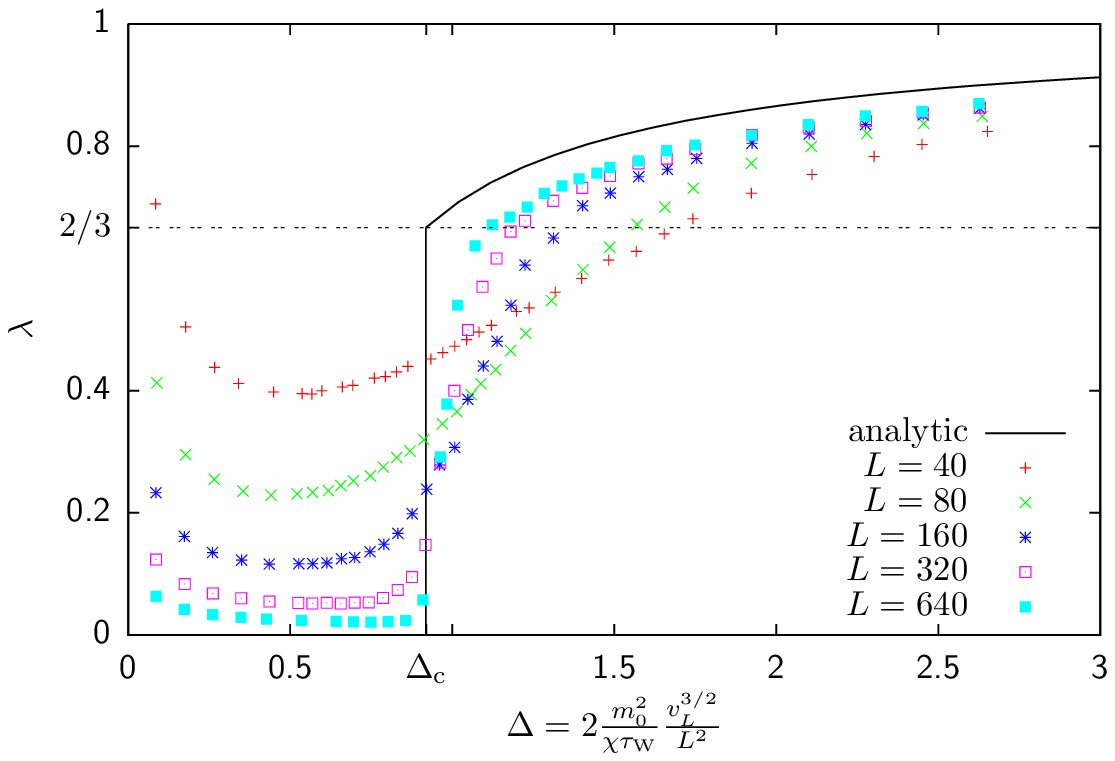}
  \caption{Fraction $\lambda$ for the two-dimensional n.n.n.\ Ising model on
square lattices of size $L=40, 80, \dots, 640$ with periodic boundary conditions
at the temperature $T=4.0 \approx 0.76 \, T_c$. The error bars are not plotted
since their size is much smaller than that of the data symbols. The  solid line
shows the analytic solution in the limit $L \to \infty$.}
  \label{fig:nnn.lambda.delta}
\end{figure}

In Figs.~\ref{fig:nn.lambda.delta} -- \ref{fig:nnn.lambda.delta} we show our
main results, the fraction $\lambda(\Delta)$ for the three observed lattices. 
The (black) solid line is the analytical value of $\lambda$ as shown in
Fig.~\ref{fig:lambda.delta.analytic}.
Clearly, for larger lattice sizes the theoretical value is approached by the
results of the simulation. Figure~\ref{fig:nn.lambda.delta} (a) shows $\lambda$
in dependence of the magnetisation $m$. In Fig.~\ref{fig:nn.lambda.delta} (b)
$\lambda$ is plotted for the same set of data points, but this time in
dependence of $\Delta$ which essentially is a rescaling with $v_L^{3/2}$. While
in (a) the important region is barely visible, the rescaling leads to a
blow up of the transition region making the theoretically predicted jump from
$\lambda_\Delta \approx 0$ to $\lambda_\Delta \approx 2/3$ at $\Delta_\mathrm{c}
\approx 0.92$ observable. This confirms that at the evaporation/condensation
transition only $2/3$ of the excess of the magnetisation goes into the droplet
while the rest remains in the background fluctuations.
The increase of $\lambda_\Delta$ for $\Delta \to 0$ can be explained
by the fact that the minimal cluster size is $1$ and not an
arbitrarily small fraction. In contrast, the excess that can be fixed
analytically using Eq.~(\ref{eq:Delta}) can be much smaller than $1$.

\begin{figure}
  \includegraphics[scale=0.7]{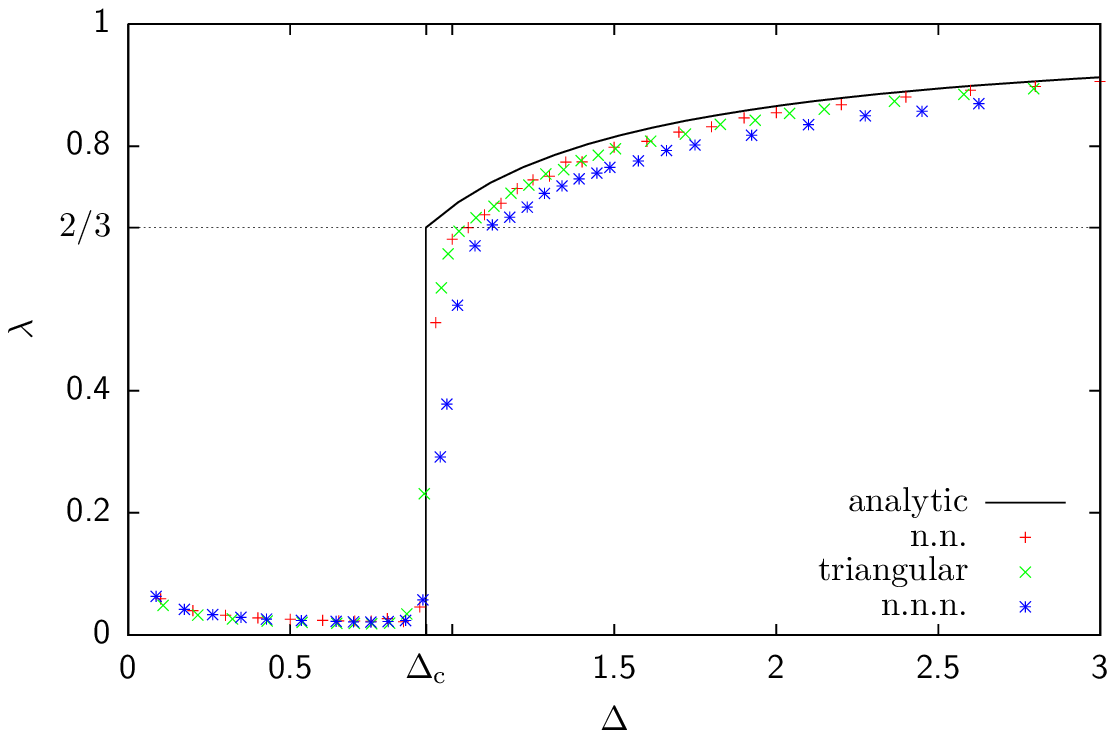}
  \caption{Comparison of the fraction $\lambda$ for the three observerd
Ising models (n.n., triangular and n.n.n.) for the size $L=640$ and the
temperatures $T=1.5, 2.4, 4.0$.}
  \label{fig:lambda.delta.640}
\end{figure}

In Fig.~\ref{fig:lambda.delta.640} we compare $\lambda$ for $L=640$ of
the three different models. The nice agreement of the data points is a clear
indication for the lattice independent universal behavior of the theory. An
explanation for the slight discrepancy between the n.n.\ and the triangular
lattice on the one side and the n.n.n.\ model on the other might be given by the
slightly different temperature ratio $T/T_\text c$ (see
Table~\ref{tab:numeric}).

\section{Conclusion}

Our Monte Carlo data clearly confirm the theoretical considerations of Biskup
{\it et al.} \cite{biskup,biskup2} for the case of the two-dimensional
next-neighbour Ising system. While their results are only valid in the
thermodynamic limit of large systems, we have shown that for practically
accessible sizes the theory can also applied. The observed finite-size scaling
behavior fits perfectly with their predictions for the infinite system.

Moreover we have demonstrated that the theory, which to date has only been
proven for the square lattice nearest-neighbour case, is actually universal in
the sense that it is independent of the underlying lattice. The Ising model on
the two-dimensional triangular lattice and on the two-dimensional next-nearest
neighour lattice both approach the theoretically expected results nicely.
Apparently, for the same relative temperature $T/T_\text c$ the finite-size
behavior is identical.

In order to achieve the correct scaling of the abscissa we presented several
methods to estimate the Wulff free energy $\tau_\text W$ numerically. While
in theory it should be straightforward to extract the value from the
distribution of the magnetisation, due to limitations in the computer time for
temperatures near the critical one, it can be more advantageous to resort to
the isotropic approximation.

All simulations were performed in thermal equilibrium and the abundance of
droplets of intermediate size could be confirmed visually by looking at the
distribution of droplets. We only state this fact here, while a more detailed
analysis and the corresponding graphs will be presented in a later publication
together with more results on the finite-size scaling behavior of the systems
and the shape of the free-energy barrier associated with the
evaporation/condensation transition.

\acknowledgments

We are indebted to Kurt Binder and Thomas Neuhaus for sharing their physical
insight into the droplet nucleation mechanism, and wish to thank Roman
Koteck\'y for helpful discussions on the formulation used in the present work.

Work supported by the Deutsche Forschungsgemeinschaft (DFG) under grants No.\
JA483/22-1 and No.\ JA483/23-1 and in part by the EU RTN-Network `ENRAGE':
``Random Geometry and Random Matrices: From Quantum Gravity to Econophysics''
under grant No.~MRTN-CT-2004-005616. Supercomputer time at NIC J{\"u}lich under
grant No.~hlz10 is also greatfully acknowledged.

\bibliography{klong}

\end{document}